\documentclass{aa}
\usepackage{natbib}
\usepackage{graphicx}
\def\xmm{{\sc Xmm--newton }}
\def\betam{$\beta$ model }
\def\deg{$^{\circ}$}
\def\msun{M$_\odot$ }

\begin{document}

\title{\object{RX~J0256.5+0006}~: a merging cluster of galaxies at z=0.36 observed with \xmm}

\author{S. Majerowicz\inst{1}, D.M. Neumann\inst{1}, A.K. Romer\inst{2}, R.C. Nichol\inst{2}, D.J. Burke\inst{3}, C.A. Collins\inst{4}}

\institute{CEA/Saclay, Service d'Astrophysique, L'Orme des Merisiers, B\^{a}t. 709, 91191 Gif--sur--Yvette Cedex, France \and Physics Department, Carnegie Mellon University, 5000 Forbes Avenue, Pittsburgh, PA15213, USA \and Harvard-Smithsonian Center for Astrophysics, 60 Garden Street, Cambridge, MA 02138, USA \and Astrophysics Research Institute, Liverpool John Moores University, Twelve Quays House, Egerton Wharf, Birkenhead L41 1LD, UK}

\offprints{Doris Neumann,\\ \email{dneumann@cea.fr}}

\date{Received  / Accepted }

\abstract{We present here a study based on \xmm data of \object{RX~J0256.5+0006}, a medium distant (z=0.36) cluster of galaxies found in the Bright SHARC catalog. The X-ray emitting intracluster medium shows a bimodal structure: one main cluster component and a substructure in the west, which very likely falls onto the main cluster centre. The subcluster shows after subtraction of the main cluster component a cometary shape pointing away from the main cluster centre, suggesting that ram pressure stripping is at work. Despite the indication of interaction between the two components we surprisingly do not find any sign of temperature gradients, which is contradictory to predictions from hydro dynamical simulations of cluster mergers.\\
Due to the non-symmetric form of the main cluster we extract three surface brightness profiles in different sectors around its centre. We see large variations between the profiles, which we quantify by \betam fitting. The corresponding $r_c$'s vary between 0.1-0.5~Mpc and the $\beta$'s between 0.5--1.2. The variations of the profiles and the \betam parameters indicate that the main cluster is not entirely relaxed. This hypothesis is strengthened further by the fact that the cluster is over luminous with respect to the (z-evolving) $L_x-T$ relation found for nearby clusters. 
\\
Galaxy clusters show a high degree of self-similarity. Comparing our profiles to the scaled reference emission measure profile of Arnaud et al. based on nearby clusters, we find that only the profile extracted north-east (NE) of the main cluster centre is similar to this reference profile. This indicates that only the NE profile is representative for the relaxed part of this cluster component. 
\\
Based on the \betam parameters of the NE profile and the spectroscopically fitted temperature of $kT=4.9^{+0.5}_{-0.4}$~keV we find for the total mass within $r_{500}$ using the hydrostatic approach $M_{500}\sim 4\times 10^{14}$\msun for the main cluster component. This value is in good agreement with the value ($M_{500}=3.9\times 10^{14}$\msun) obtained using the z-evolving $M_{500}-T$ relation from the HIFLUGCS sample based on nearby clusters. A non-z-evolving $M-T$ relation is only marginally consistent with our result. This is an indication that there exists evolution in the $M-T$ relation, as predicted from simple scaling laws. Calculating the corresponding gas mass fraction we find $f_g\sim 18-20\%$ which is in good agreement with other work. 
\\
We also develop a simple on-axis merger model for the cluster. As input we use the projected distance of the subcluster to the main cluster centre and the velocity difference of main and subcluster based on four galaxy redshifts spectroscopically measured with the Kitt Peak telescope. Together with a simple ram pressure model we find that the most likely physical distance of the subcluster to the main cluster lies between $0.6<d<1.0$~Mpc. The coherent results of on-axis merger and ram pressure model suggest that the merger in this cluster is indeed on-axis and not an off-axis merger with a large impact parameter.
\\
We find for the ratio of subcluster to main cluster mass values between 20--30\% which indicates that the merger in \object{RX~J0256.5+0006} is a major merging event.}

\authorrunning{S. Majerowicz et al.}
\titlerunning{RX~J0256.5+0006~: a merging cluster at z=0.36}

\maketitle
\keywords{galaxies: clusters: individual: RX~J0256.5+0006 -- galaxies: clusters: general -- cosmology: observations}

\section{Introduction}
According to the scenario of hierarchical structure formation, large-scale structures like galaxy clusters form through the accretion of smaller units. Clusters of galaxies are still forming today via merger events (Kempner, Sarazin \& Ricker 2002~; Neumann et al. 2003). During cluster mergers, the hot X-ray emitting intracluster medium (ICM) is heated locally by shock waves and compression. X-ray observations of the ICM allow to study the dynamical state of clusters of galaxies. The spectro-imaging capabilities of {\sc Xmm-newton} and {\sc Chandra} allow studies with unprecedented precision of the ICM in interacting clusters up to relatively high redshift (Markevitch \& Vikhlinin 2001~; Worall \& Birkinshaw 2003~; De Filippis, Schindler \& Castillo-Morales 2003). These observations contribute to the question on how large scale structure and galaxy clusters form and grow, which is connected to the evolution of internal physical cluster properties as well as cosmology.

We present in this paper the EPIC \xmm observations of \object{RX~J0256.5+0006}, which is a medium redshift (z=0.36) cluster of galaxies first detected in {\sc rosat} {\sc pspc} data and is a member of the Bright {\sc sharc} --- Serendipitous High-redshift {\sc Rosat} Cluster Survey --- cluster catalog (Burke et al. 1997~; Collins et al. 1997~; Romer et al. 2000~; Burke et al. 2003). This cluster is one of several {\sc sharc} clusters selected for X-ray follow-up by {\sc Xmm--Newton} (see Arnaud et al. 2002b~; Majerowicz, Arnaud \& Neumann (2002)~; Lumb et al. 2003).

The paper is organized as follows: in section \ref{data}, we describe the observation and the data treatment. In section \ref{morph}, we present our imaging analysis. Our study shows that \object{RX~J0256.5+0006} is most likely composed by two cluster components, one main cluster component and a smaller subcluster West of the main cluster centre. Sec.\ref{sbp} focuses subsequently on the surface brightness profile of the main cluster. This is followed by section \ref{specimag}, in which we present a spectro-imaging analysis which comprises a search for temperature variations via the construction of a hardness ratio map as well as spectral fits. We conduct a spectral analysis in distinct cluster regions, selected in function of the likely merger scenario found in this particular cluster. In section \ref{mcprop} we discuss the main cluster properties. This is followed by section \ref{optical}, in which we present optical observations of \object{RX~J0256.5+0006} and compare them to \xmm data. Using the previously obtained results, the dynamics of the merging event is modeled in section \ref{dyncluster}. This allows us to put constraints on the merger geometry and the physical distance between the two cluster components. We also give estimates on the luminosity and mass of the subcluster. The discussion and conclusions of all our results are given in section \ref{conclusion}.

In this paper, we use a flat and low density cosmology with H$_{0}$=50\,km/s/Mpc, $\Omega_{m}$=0.3 and $\Omega_{\Lambda}$=0.7. In this cosmology, at the redshift of the cluster (z=0.36), one arc minute corresponds to 423\,kpc. Uncertainties and errors are 90\% level unless specified otherwise.

\section{Data treatment}\label{data}

\subsection{Observation}

\object{RX~J0256.5+0006} was observed with \xmm  during its 217-th revolution for a total of 25.3\,ks. The calibrated event lists were retrieved from the Science Operation Centre. We concentrate in this article on EPIC data. The {\sc thin}1 filter was used during the observation. The observation mode for EMOS and EPN cameras was Full Frame and Extended Full Frame Mode, respectively.

Galaxy clusters are extended X-ray sources with relatively low surface brightness emission. Thus contamination of background emission can play an important role, especially in the outskirts of clusters. Therefore proper background subtraction is crucial to obtain reliable results. As background, we use in the following blank sky observations (one for each camera) provided by Lumb (2002)\footnote{The background event files can be down-loaded from \textsf{ftp://xmm.vilspa.esa.es/pub/ccf/constituents/extras/background/}}. The corresponding files consist of several high galactic latitude exposures performed by \xmm. To avoid contamination, bright point sources present in these observations are cut out. 

For our analysis, we only select single ({\sc pattern} 0) events, for the EPN data and {\sc pattern} 0 to 12 events for EMOS data sets. This selection was performed for the source and background observations.

\subsection{Vignetting correction}\label{vign}

The sensitivity of the EPIC cameras is not constant but decreases with increasing off-axis angle. To correct for this effect, called vignetting, we use the method described in Arnaud et al. (2001). In this method each detected event obtains an individual weight factor which is the ratio of the central effective area at the event energy over the effective area at the energy and location of the event in the camera. Therefore when creating images, spectra or surface brightness profiles, the event weight factors are added up into bins instead of the individual event counts themselves. Thus, a proper correction of the vignetting is assured. The error propagation takes into account the fact that we use weight factors and not photons.

\subsection{Background treatment}

\subsubsection{Flare rejection}

\begin{figure}
\resizebox{\hsize}{!}{\rotatebox{-90}{\includegraphics{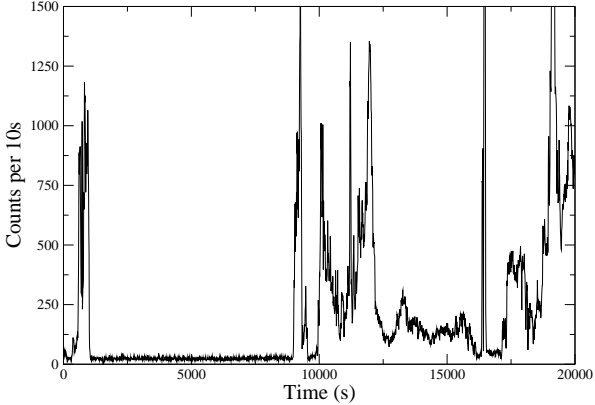}}}
\caption{Light curve of the \object{RX~J0256.5+0006} observation for the EPN camera. Only events with an energy greater than 300\,eV are selected and subsequently binned in time intervals of 10\,s.}\label{lc}
\end{figure}

The background of \xmm shows sometimes periods of high intensity which are the result of soft protons. These periods are also called flares. To illustrate these flares the light curve of the \object{RX~J0256.5+0006} observation is displayed in Fig.\ref{lc}. Since these periods of high intensity limit severely the signal-to-noise of the observation, we discard them without further study.

For the flare rejection, we use the method described in Majerowicz, Neumann \& Reiprich (2002) i.e. we only keep time intervals where there are less than 15 counts per 100\,s in the 10 to 12\,keV energy band for each EMOS camera and 22 counts per 100\,s in the 12 to 14\,keV energy band for EPN (please note we only use Pattern=0 events for EPN). The remaining exposure times are 10550\,s for EMOS1, 10340\,s for EMOS2 and 7100\,s for EPN.

\subsection{Remaining background components}\label{back}

Beside the particle background there exist two other kinds of background: a background induced by highly energetic particles which is relatively constant with time and is well described by our used blank sky observations, and the cosmic X-ray background (hereafter {\sc cxb}) which is dependent on the sky position (see Snowden et al. 1997).

To correct for these different background components, we use the method described in Majerowicz, Neumann \& Reiprich (2002) and Pratt, Arnaud \& Aghanim (2001). We outline it briefly in the following.

The intensity of the high energy particle background varies typically within 5 or 10\,\% from observation to observation. In order to account for these variations, we normalize the background at energies above 10\,keV where the emission is dominated by these particles. To calculate the normalization factor for the EMOS cameras we divide the count rate in the 10 to 12 keV band of the source observation by the count rate in the same energy band of the blank sky observation. Since the EPN camera has a higher quantum efficiency at high energies we use for the EPN-data the 12 to 14 keV energy band instead of the 10 to 12 keV band.

To correct for the local {\sc cxb} which is not necessarily well represented by the blank sky observations, we perform a double background subtraction: first, the blank sky observations are subtracted from regions inside and outside of the cluster. The latter represents the local background. For the subtraction we use the normalization factor calculated from high energy events  mentioned above. Then, the remaining residuals of the region outside the cluster are subtracted from the blank sky subtracted regions in the cluster. The result is thus corrected for all the different background components.

This method will be described with more detail in the case of spectra (Sec.\ref{temp}), surface brightness profiles (Sec.\ref{sbp}) and images (Sec.\ref{hrmap}).

In order to avoid source confusion 46 clearly detected point sources are excluded from our analysis.

\section{Cluster Morphology}\label{morph}

\begin{figure}
\vspace*{5cm}
\caption{Combined EMOS1, EMOS2 and EPN image of \object{RX~J0256.5+0006} in the 0.3 to 2.0\,keV energy band. The image is convolved with a Gaussian ($\sigma$=6.6'') filter. The black contours are the residuals after the subtraction of the elliptical \betam from the data (see section \ref{morph}). The contours are expressed in terms of statistical significance (the lowest contour is 3\,$\sigma$ and the step width is 1\,$\sigma$). The center of the best fit elliptical $\beta$ model is represented by the black cross.}\label{signif}
\end{figure}

Fig.\ref{signif} shows the \xmm image of \object{RX~J0256.5+0006} in the 0.3 to 2.0\,keV energy band. The cluster exhibits a clear bimodal structure. An important question is whether the two components which are responsible for the bimodality are physically connected and in interaction or whether they are merely due to a chance alignment. We will address this issue in the following.

The projected distance $d_{\mathrm{min}}$ of the two components is about 0.8' which corresponds to a physical distance of 350\,kpc at the redshift of \object{RX~J0256.5+0006} with our adopted cosmological parameters.

Since the cluster shows an elongation in the North South direction we model the hot gas distribution of the main component, with an elliptical \betam
(see also Neumann \& B\"{o}hringer 1997). The elliptical \betam has the following form~:
\begin{equation}
S(x,y) = S_{0}{(1+F_{1}+F_{2})}^{-3\beta+1/2}+\mathrm{B}
\end{equation}
with~:
\begin{eqnarray}
F_{1}  & = & {{{((x-x_{c})\cos\alpha + (y-y_{c})\sin\alpha)}^{2}}\over{{M}^{2}}}\nonumber
\\
F_{2}  & = & {{{(-(x-x_{c})\sin\alpha + (y-y_{c})\cos\alpha))}^{2}}\over{{m}^{2}}}\nonumber
\end{eqnarray}
where $(x_{c},y_{c})$ are the coordinates of the cluster center, $M$ is the major axis, $m$ the minor axis, and $\alpha$ the position angle. The background level is included via the parameter $\mathrm{B}$.

The fitting technique relies on Gaussian statistics via the $\chi^{2}$ test. However, in the outer regions of the image, the number of photons per pixel is low and more appropriately described by Poisson statistics. To correct for this effect, we apply to the image a Gauss filter with a $\sigma$ of 6.6'' (see also Neumann 1999 for more detail). We also exclude point sources in the image and the western structure. The best fit parameters are listed in Tab.\ref{ellbeta}. The ratio $m/M\sim0.80$ confirms the apparent ellipticity of the main component.

Since the emission from the western component is contaminated by the main component emission, we subtract the elliptical \betam best fit (see Tab.\ref{ellbeta}) from the data in order to determine the morphology of this substructure. The residuals obtained in terms of significance are shown in Fig.\ref{signif}.

\begin{table}
\begin{center}
\begin{tabular}{ll}
\hline
\hline
Parameter & best fit
\\
\hline
$\beta$ & 0.83
\\
$M$ & 370\,kpc
\\
$m$ & 300\,kpc
\\
$\alpha$ & 95\deg
\\
$x_{c}$ & 2$^{\mathrm{h}}$\,56$^{\mathrm{m}}$\,34.3$^{\mathrm{s}}$
\\
$y_{c}$ & 00\deg\,06'\,11''
\\
\hline
\end{tabular}
\end{center}
\caption{Elliptical \betam best fit parameters of the main component of \object{RX~J0256.5+0006}. These values are obtained by fitting the X-ray emission distribution of the main cluster by excluding the western substructure.}\label{ellbeta}
\end{table}

The western structure is very significant at a detection limit which largely exceeds 3\,$\sigma$. The residuals can be relatively well described by an ellipse with major and minor axes of 1.7' and 1.4', respectively. At the redshift of the cluster, this corresponds to 720\,kpc\,$\times$\,590\,kpc. These values indicate that this structure is very likely a small galaxy cluster. Henceforth, we will refer to this western component as the `` subcluster ''.
The morphology of this subcluster resembles somewhat a comet and indicates strongly that this object is in interaction with the main cluster~: the elongated structure suggests that gas is pushed out of the subcluster due to ram pressure stripping as it encounters the ICM of the main cluster.

To examine the robustness of the comet-like shape of the structure, we perform two different tests~:
\begin{description}
\item[-]we vary the $\beta$ value between 0.75 and 0.9 for the elliptical $\beta$ model~;
\item[-]we subtract a spherical $\beta$ model instead of the elliptical model. Two spherical models are used, the first with a core radius of $m$, the second with a core radius of $M$.
\end{description}
In each case, the particular comet-like shape of the subcluster remains.

The comet-like structure, indicating that ram pressure stripping is acting on the subcluster, suggests that the subcluster is indeed interacting with the main cluster centre. We will discuss the distance between main cluster and subcluster in more detail in Sec.\ref{dyncluster}.

Numerical simulations (e.g. Ricker \& Sarazin 2001) show that during a merger event of two clusters, the hot gas of the smaller cluster passes through the main cluster core once then goes back to the core to be absorbed. In this frame, the orientation of the \object{RX~J0256.5+0006} subcluster isophots indicates that this cluster has not yet passed through the main cluster center. Markevitch et al. (2002) found for the cluster \object{1E0657-56} a ``~bullet~'' structure comparable to the comet-like structure of \object{RX~J0256.5+0006}. However, in the case of \mbox{1E0657-56}, the orientation of the isophots revealed that the ``~bullet~'' already traversed the cluster core.

Additionally, the residual map also shows significant residuals inside the main cluster region. In particular, an offset is detected between the maximum of X-ray emission and the center of the best fit elliptical \betam (see Fig.\ref{signif}, black cross and northern excess). This will be more discussed in Sec.{sbp} Sec.\ref{galpos}.

\section{Surface brightness profiles of the main component}\label{sbp}

We extract vignetting corrected surface brightness profiles of the main cluster. We only select events in the 0.3 to 3.0\,keV energy band to optimize the signal-to-noise ratio. We discard regions where point sources are present and we apply a binning of 3.3'' which is the width of each concentric annulus.

One surface brightness profile is produced for each EPIC camera and the so obtained profiles are then subsequently summed in one single profile. We also extract surface brightness profiles from the blank sky observations with the same detector coordinates. These background profiles are subtracted from the cluster profiles. The remaining background emission is due to the sky variations of the {\sc cxb}. To correct for this emission, we estimate the remaining mean surface brightness value in the annulus between 4' and 7' which is well beyond the detected cluster emission and we subtract this value from the profile. We finally group the data into bins with i) a signal-to-noise ratio of at least 3$\sigma$ above background and ii) at the same time a logarithmic binning in which the binsize at radius $\theta$ is: $\Delta \theta \geq 1.15\times \theta$. This kind of  binning is similar to the binning presented in Arnaud, Aghanim \& Neumann (2002a). 

In Sec.\ref{morph} it was shown that there is an offset between the fitted center of the main cluster and the maximum of the X-ray emission of roughly 8''. This offset corresponds to roughly 60~kpc. The location of the maximum of the emission coincides with the location of the main cluster galaxy and might be contaminated by a point source such as a central AGN. However, in order to obtain monotonically decreasing surface brightness profiles, we choose the X-ray maximum of the cluster as the profile extraction center. We evaluate the potential error based on this choice of center by comparing the subsequent 1d-fit results with the 2d-fit results, which possess different \betam centres: the $\beta$ from the 2d-fit ($\beta=0.83$ -- please note: we do not estimate the uncertainty of this parameter in our 2d-fit) lies within the found $\beta$-values of the 1d-fit of the overall profile with and without the extraction of the central excess emission (see Tab.1 and Tab.2). Furthermore the geometrical mean of the minor and major axis core radius of the 2d-fit (0.3~Mpc), is similar to the core radius found for the 1d-profile without cutting out the central excess region. We conclude therefore that the choice of the center for extracting the surface brightness profile is not important and that the potentially existing point source in the cluster center does not influence significantly our fit results.

\begin{figure}
\resizebox{\hsize}{!}{\includegraphics{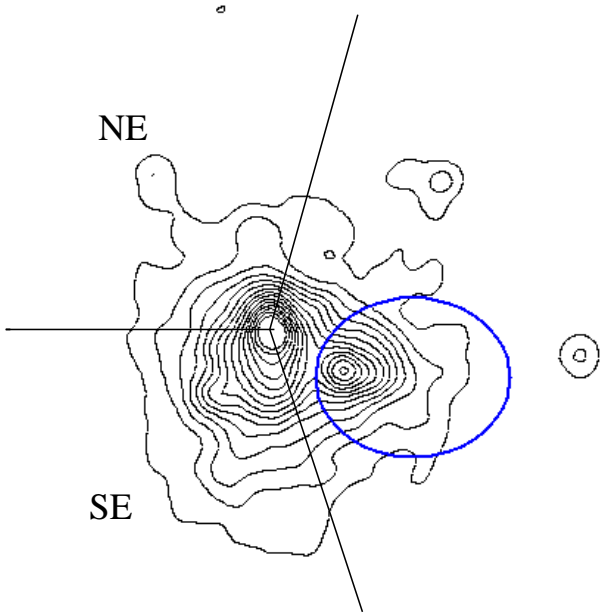}}
\caption{X-ray emission contours from the 0.3 to 2.0\,keV image of \object{RX~J0256.5+0006} (Fig.\ref{signif}) and the surface brightness extraction areas where the blue ellipse region is excluded.}\label{sbpext}
\end{figure}

\begin{figure}
\resizebox{\hsize}{!}{\includegraphics{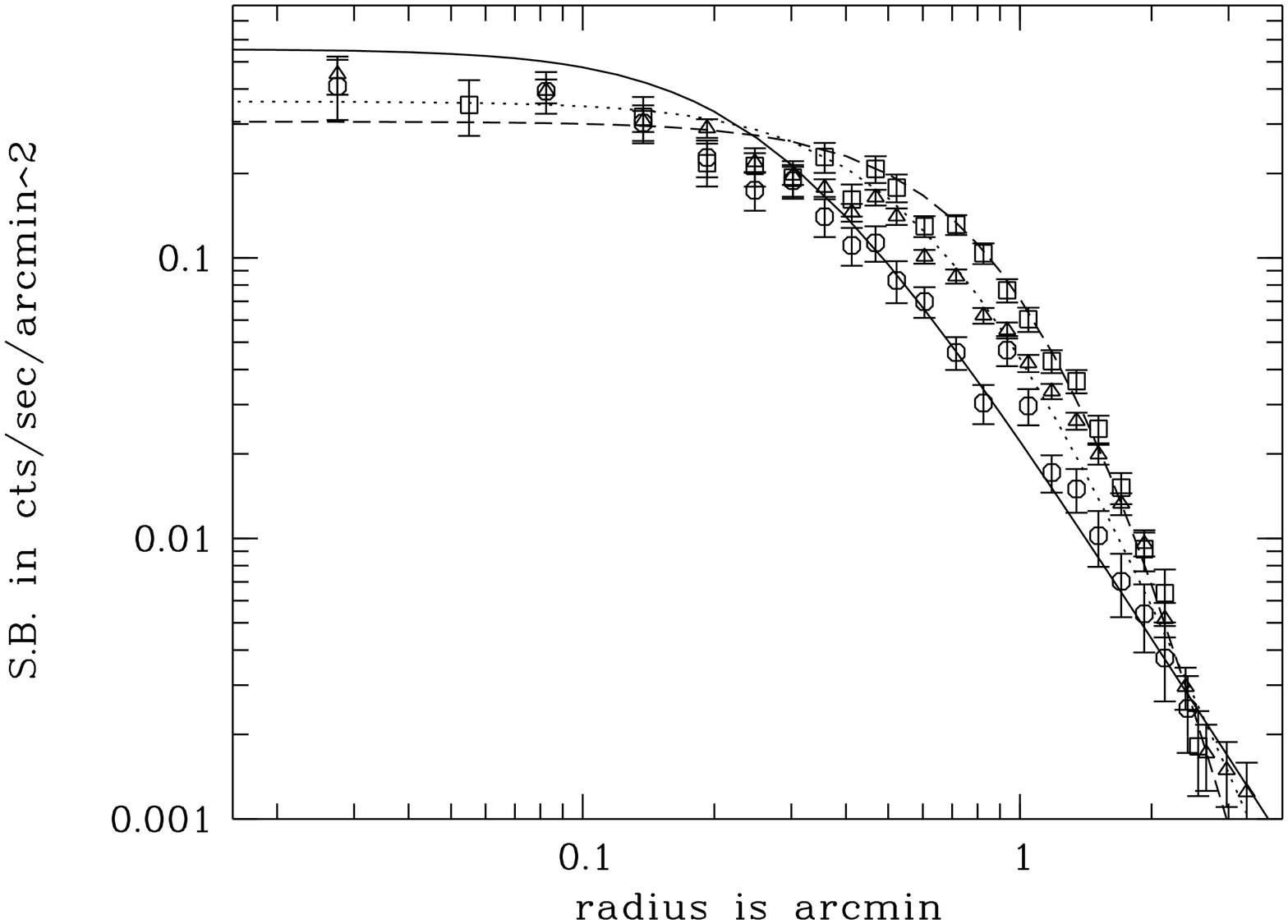}}
\caption{Background corrected surface brightness profiles from the main component of \object{RX~J0256.5+0006} in the 0.3 to 3.0\,keV energy band (the three {\sc epic} camera, EMOS1, EMOS2 and EPN are added). The different regions chosen for the extraction are shown in Fig.\ref{sbpext}. The triangles show the overall profile without angular selection, the circles correspond to the NE area and the red boxes to the SE area. The lines show the best fit \betam profiles: full line for the NE-profile, dotted line for the overall profile and dashed line for the SE profile. Please note: the fits take into account PSF-effects and therefore the fitted profile of the NE profile lies above the data points in the innermost region.}\label{sbpcomp}
\end{figure}

Fig.\ref{sbpext} shows the X-ray contours of the cluster. As one can see the contours in the North-Eastern (NE) part of the cluster are more compressed than in the South-Eastern (SE) part. Because of this we extract surface brightness profiles in different regions, which are defined in Fig.\ref{sbpext}: the first region covers the whole area with the exception of an ellipse defined in Fig.\ref{sbpext} surrounding the subcluster. The second region only represents the NE part of the cluster and the third profile is selected in the SE cluster part. The resulting profiles are shown in Fig.\ref{sbpcomp}.

The overall cluster profile can be detected out to a radius of 3.5'. This corresponds to a physical radius of roughly 1.5\,Mpc. Even with our adopted logarithimic binning (see above) it is not possible to obtain the chosen signal-to-noise ratio (3$\sigma$) for radii greater than this value.

Not surprisingly the NE and SE profiles show very different shapes. The SE profile is much steeper than the NE profile at large radii and the core of the SE profile is much larger than the NE core. This  is equivalent to the fact that the contours in the NE part of the cluster are more compressed than in its SE part. In order to quantify the differences, we fit a \betam to the different surface brightness profiles.

\begin{table*}
\begin{center}
\begin{tabular}{ccccccc}
\hline
\hline
 & \multicolumn{6}{c}{Radial profiles}
\\
Parameters & \multicolumn{2}{c}{Overall} & \multicolumn{2}{c}{NE} & \multicolumn{2}{c}{SE}
\\
 & $a$ & $b$ & $a$ & $b$ & $a$ & $b$
\\
\hline
$\chi^{2}_{\mathrm{red}}$ ($dof$) & 3.3 (22) & 2.0 (17) & 1.2 (19) & 1.2 (14) & 0.9 (17) & 0.9 (12)
\\
n$_{\mathrm{e}0}$ (10$^{-3}$\,cm$^{-3}$) & 5.23$^{+0.20}_{-0.18}$ & 4.20$^{+0.16}_{-0.15}$ & 9.02$^{+0.98}_{-0.90}$ & 7.02$^{+1.91}_{-1.01}$ & 4.05$^{+0.21}_{-0.17}$ & 4.05$^{+0.34}_{-0.16}$
\\
r$_{\mathrm{c}}$ (kpc) & 286$^{+37}_{-33}$ & 378$^{+59}_{-50}$ & 117$^{+40}_{-30}$ & 153$^{+85}_{-68}$ & 545$^{+210}_{-128}$ & 546$^{+231}_{-148}$
\\
$\beta$ & 0.773$^{+0.059}_{-0.049}$ & 0.891$^{+0.099}_{-0.079}$ & 0.572$^{+0.063}_{-0.047}$ & 0.611$^{+0.114}_{-0.073}$ & 1.21$^{+0.60}_{-0.26}$ & 1.22$^{+0.74}_{-0.30}$
\\
\hline
\end{tabular}
\end{center}
\caption{Best fit $\beta$ model parameters for the different extracted surface brightness profiles of the main cluster. $a$: the fit is performed over the whole radial profile. $b$: the first five bins (corresponding to 120~kpc or 17$''$-- the central excess emission) are excluded for the fit (see section\ref{morph}).}\label{betasbp}
\end{table*}

The $\beta$ model gives the relationship of the electron density as function of radius (Cavaliere \& Fusco-Femiano 1976)~:
\begin{equation}
\mathrm{n}_{\mathrm{e}}(r)=\mathrm{n}_{\mathrm{e}0}{\Bigg(1+{\bigg({{r}\over{r_{\mathrm{c}}}}\bigg)}^{2}\Bigg)}^{-{{3}\over{2}}\beta}\label{ne}
\end{equation}
in which the free parameters are n$_{\mathrm{e}0}$, the central electron density, $r_{\mathrm{c}}$, the core radius and $\beta$ a slope parameter. The surface brightness $S_{x}$ for the $\beta$ model can be written as~:
\begin{equation}
S_x(r)=S_0{\Bigg(1+{\bigg({{r}\over{r_c}}\bigg)}^{2}\Bigg)}^{-3\beta+{{1}\over{2}}}\label{sb}
\end{equation}
where the $S_{\mathrm{x}0}$ is the central brightness.

We fit the surface brightness profiles with a $\beta$ model convolved with the Point Spread Function (hereafter PSF) of the detectors (see Ghizzardi 2001; Arnaud et al. 2002b). We perform two $\beta$ model fits for each radial profile: the first by using all the bins and the second by excluding the first five bins to avoid the influence of the detected excess, which might be linked to a point source (Sec.\ref{morph}). From the best fit parameters including their errors and the mean temperature of the main cluster (see Sec.\ref{temp}) of 4.9$^{+0.5}_{-0.4}$\,keV, $n_{\mathrm{e}0}$ can be determined.The results are shown in Tab.\ref{betasbp} and illustrated in Fig.\ref{sbpcomp}.

Not surprisingly, the best fit parameters are different for the different profiles. Particularly, one can note that the best fit $\beta$ for the SE profile is very high. The fit values also change when the central part of the profile is cut out by giving generally larger $\beta$ values. However, the different fit parameters found for the entire profile and for the same profile with the central region cut out agree generally within the error bars. 

\section{Spectro-imaging analysis}\label{specimag}

\subsection{Hardness ratio map}\label{hrmap}

There are strong indications that \object{RX~J0256.5+0006} is in a merger state. In this case it is very likely to observe temperature variations in the ICM. To visualize these possible variations we construct a hardness ratio image of the cluster, which consists of diving a background subtracted image in the hard energy band by a background subtracted image in the soft photon band. In order to ensure an optimal choice of the energy bands for the hardness ratio image we simulate an \xmm spectrum with galactic absorption n$_\mathrm{H}$=6.6$\times$10$^{20}$\,cm$^{-2}$ (Dickey \& Lockman 1990), $z=0.36$ and $kT=4.9$~keV (see also below). Based on this spectrum we choose for the soft band 0.3-1.3\,keV and for the hard band 1.3-7.0\,keV. With this choice we ensure to have the same number of photons in each band, which minimizes the error in the resulting hardness ratio map. 

Since the cluster has low surface brightness and in order to have sufficient counts in each image pixel to obtain statistical significant results for the hardness ratio, we need to adopt a large pixel size. From the previous fits (see Sec.\ref{morph} and \ref{sbp}), we compute the total count rate coming from the cluster inside a radius of 2.5' in the hard band and we find about 0.1\,cts/s. By assuming that the emission is flat, we estimate a pixel size of 14.3''\,$\times$\,14.3'' in order to have at least 3 detected cluster photons in each pixel of the hard energy image.

To correct for the background, we also extract images in the same energy bands from the blank sky observations. We subtract the background  from the cluster images according to the normalization defined in Sec.\ref{back}. To correct for the {\sc cxb} contribution, we first extract surface brightness profiles in the soft and hard energy band for each camera. Then, we subtract the profiles of the blank sky observations from the cluster profiles in the same energy band. The residuals at large radii (outside the cluster emission region) give the correction for the local CXB contribution which is assumed to be constant across the entire field-of-view. We subtract the obtained averaged local CXB contribution from the blank sky corrected image pixels in the soft and hard energy band. This background correction is performed for each camera image. The images are subsequently summed to obtain a single image in each energy band.

\begin{figure}
\vspace*{5cm}
\caption{Hardness ratio map of \object{RX~J0256.5+0006} obtained by dividing the background corrected image in the 1.3-7.0\,keV energy band by the background corrected image in the 0.3-1.3\,keV band. Before the division the two images were smoothed by a Gauss filter with $\sigma$=14.3''. In the region in which cluster emission is found, the temperatures vary between 4.0 and 5.5\,keV (blue colors correspond to lower temperatures than green colors.). The contours of the substructure (see Sec.\ref{morph}) are overlayed in black.}\label{hrmapimg}
\end{figure}

All our cluster images are thus fully background corrected. To avoid too strong statistical fluctuations from pixel to pixel, we finally apply a Gauss filter with $\sigma$=14.3'' before the division of the hard energy by the soft energy image. The final hardness ratio map is presented in Fig.\ref{hrmapimg}.

Converting the hardness ratio into temperature gives values for each pixel which vary between 4.0\,keV and 5.5\,keV. This excludes the two pixels in North-West which show an extremely high hardness ratio (corresponding to temperatures greater than 10\,keV in the case of bremsstrahlung). These pixels correspond to a point-like source. 

The hardness ratio map does not show any sign of significant temperature gradients in the cluster. This result is surprising for the area between the two interacting clusters in which heating due to merging is expected.

\subsection{Temperature estimates}\label{temp}

To verify the results given by the hardness ratio map presented in Sec.\ref{hrmap}, we perform spectral fits in order to obtain more reliable temperature estimates. 

Since we apply the described vignetting correction method (see Sec.\ref{vign}), it is possible to use directly the on-axis response matrices~: \textsf{m1{\_}thin1v9q20t5r6{\_}all{\_}15.rsp} (EMOS1), \textsf{m2{\_}thin1v9q20t5r6{\_}all{\_}15.rsp} (EMOS2) and \textsf{epn{\_}e\mbox{}f20{\_}sY9{\_}thin.rsp} (EPN) for the spectral fitting. To avoid remaining uncertainties at very low energy, we exclude for our analysis events below 0.3\,keV.

\begin{figure}
\resizebox{\hsize}{!}{\includegraphics{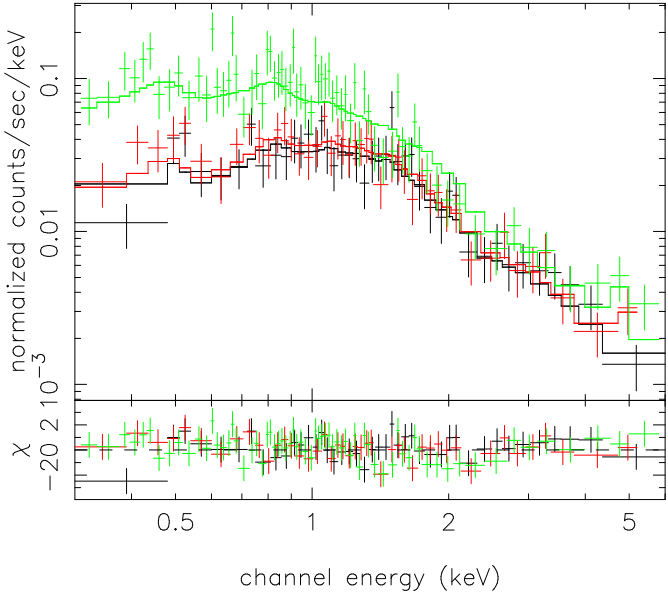}}
\caption{EMOS1 (black), EMOS2 (red) and EPN (green) spectra of \object{RX~J0256.5+0006} in the region 2 (see also Fig.\ref{tempimg}). The presented spectra are vignetting and background corrected (see sections \ref{vign} and \ref{back}).}\label{spectre}
\end{figure}

We extract spectra in different regions (see Fig.\ref{tempimg}) and correct for background (see Sec.\ref{back}. We estimate the {\sc cxb} contribution from an annulus between 4.5' and 11'. We define the following regions: the main cluster center (T$_{1}$), the outer parts of the main cluster (T$_{2}$), the subcluster region (T$_{3}$) and the area in front of the subcluster region pointing towards the main cluster centre (T$_{4}$).

We group the background subtracted spectra such that each bin has a signal-to-noise ratio greater than 3$\sigma$ after background subtraction.

\begin{table}
\begin{center}
\begin{tabular}{cccc}
\hline
\hline
Region & T (keV) & A (solar) & $\chi^{2}_{\mathrm{red}}$ ($dof$)
\\
\hline
1 & 4.9$^{+0.6}_{-0.5}$ & 0.36$^{+0.23}_{-0.21}$ & 1.1 (203)
\\
2 & 4.7$^{+0.8}_{-0.6}$ & 0.43$^{+0.33}_{-0.29}$ & 0.9 (154)
\\
1+2 & 4.9$^{+0.5}_{-0.4}$ & 0.36$^{+0.18}_{-0.17}$ & 1.0 (322)
\\
3 & 5.7$^{+1.3}_{-1.0}$ & 0.30 frozen & 0.9 (108)
\\
4 & 5.2$^{+2.0}_{-1.2}$ & 0.30 frozen & 0.8 (52)
\\
\hline
\end{tabular}
\end{center}
\caption{Results from the isothermal model fits of the combined EMOS1, EMOS2 and EPN spectra in the regions defined in Fig.\ref{tempimg}.}\label{specfit}
\end{table}

We use {\sc Xspec} (Arnaud 1996) to fit the resulting spectra with an absorbed and redshifted isothermal model ({\sc wabs$\times$mekal}, see Morrisson \& McCammon 1983, Mewe et al. 1985, Kaastra 1992, Liedahl et al. 1995) with an absorption fixed to the galactic value of n$_\mathrm{H}$=6.6$\times$10$^{20}$\,cm$^{-2}$ (Dickey \& Lockman 1990) and a redshift fixed to 0.36. The resulting best fit parameters are presented in Tab.\ref{specfit} and Fig.\ref{tempimg}. To illustrate the quality of the spectra, we show in Fig.\ref{spectre} the spectra used for the estimate of T$_{2}$.

\begin{figure}
\vspace*{5cm}
\caption{X-ray emission image from the 0.3 to 2.0\,keV image of \object{RX~J0256.5+0006} (Fig.\ref{signif}) and temperature estimates in our selected cluster regions.}\label{tempimg}
\end{figure}

The fact that T$_{1}$ and T$_{2}$ are similar indicates that the main cluster does not host a cooling flow. Furthermore, we confirm with our temperature estimates (see Sec.\ref{hrmap}) the absence of strong temperature variations. The ICM appears fairly isothermal. We will discuss this in more detail in Sec.\ref{dyncluster}.

To check for the proper estimation of the redshift we leave this parameter as a free fit parameter for region 1+2. The corresponding fitted redshift is 0.351$^{+0.018}_{-0.023}$ which is in agreement with the redshift determined from optical spectra (see Romer et al. 2000 and Sec.\ref{optical}). Since there might exist a large velocity difference between the main cluster and the subcluster we also perform another fit leaving the redshift as a free fit parameter in region 3. Unfortunately, the lack of statistics does not allow to constrain the redshift in this region.

\section{Main cluster properties}\label{mcprop}

\subsection{Luminosity estimates}\label{lumin}

Due to the pressence of background, we do not detect the cluster emission up to the virial radius. It is thus possible that we understimate the total count rate of the cluster since we miss out on the external parts of the cluster. To overcome this potential problem we use the elliptical  \betam fit parameters (presented in Tab.\ref{ellbeta}, based on photons in the energy band 0.3-2.0~keV) for the cluster emission, which we extrapolate up to the virial radius $r_v$, which is defined below. We calculate the corresponding count rate of the model extrapolated to $r_v$ and convert it into a bolo-metric luminosity estimate.

The case of evolution of galaxy cluster properties with redshift is still an open issue since many studies give opposite results (e.g. Vikhlinin et al. 2002~; Borgani et al. 2001). Since \object{RX~J0256.5+0006} is not a nearby cluster, we choose to apply an evolution term according to the self-similar model (Bryan \& Norman 1998~; Eke, Navarro \& Frenk 1998) to all the scaling laws used in this paper. According to the spherical collapse model, the virial radius $r_{v}$ is~:
\begin{eqnarray}
r_{v} & = & 3894\,{\Delta_{z}}^{-{{1}\over{2}}}{(1+z)}^{-{{3}\over{2}}}{\bigg({{T}\over{10\,\mathrm{keV}}}\bigg)}^{{{1}\over{2}}}\ \mathrm{kpc}\nonumber\\
\mathrm{with}\ \Delta_{z} & = & {{\Delta_{c}\Omega_{m}}\over{18\pi^{2}\Omega_{z}}}\ \mathrm{and}\ \Omega_{z}={{\Omega_{m}(1+z)^{3}}\over{\Omega_{m}(1+z)^{3}+\Omega_{\Lambda}}}\label{evorv}
\end{eqnarray}
where $\Delta_{c}$ is the cluster density contrast (see Bryan \& Norman 1998) and $\Omega_{z}$ the density parameter of the Universe (for more detail see Arnaud, Aghanim \& Neumann 2002a). We find $\Delta_{z}=0.42$ for our adopted cosmology and thus $r_{v}$=2650\,kpc. This corresponds to 6.3' and confirms that we do not detect the main cluster of \object{RX~J0256.5+0006} up to the virial radius but only up to 0.55$\times r_{v}$ (Sec.\ref{sbp}). However, the missing contribution beyond the detection radius is estimated to be only about 3\,\% which is similar to the statistical uncertainties.

For the conversion into luminosity, we use a cluster temperature of 4.9$^{+0.5}_{-0.4}$\,keV (see Sec.\ref{temp}). We obtain a bolo-metric luminosity
of $L_{bol,x}=1.49\pm0.10\times10^{45}$\,ergs/s.

If we apply the measured $L_x$-T relation found by Arnaud \& Evrard (1999) by assuming a standard evolution term according to the self-similar model~:
\begin{equation}
L_x=\frac{\Delta_{z}^{1/2}}{\Delta_{z=0}^{1/2}}{(1+z)}^{{{3}\over{2}}}10^{45.06\pm0.03}\,\frac{\mbox{erg}}{\mbox{s}}\bigg({T\over\mathrm{6\,keV}}\bigg)^{2.88\pm0.15}\label{lxt}
\end{equation}
the measured temperature of 4.9$^{+0.5}_{-0.4}$\,keV gives a luminosity of 8.7$^{+2.8}_{-2.3}\times10^{44}$\,ergs/s (the uncertainties are only based on temperature uncertainties, since they are much larger than the uncertainties based on the photon number.). According to equ.\,(\ref{lxt}), a temperature of about 5.9\,keV is needed to reproduce a luminosity of 1.5$\times10^{45}$\,ergs/s. This temperature exceeds the temperature measurements including errors by 0.5~keV. 
This discrepancy of observed and predicted luminosity suggests that this cluster is not entirely in equilibrium.

\subsection{The scaled emission measure profiles of the main cluster}

Arnaud, Aghanim \& Neumann (2002a) found that scaled emission measure profiles of distant clusters are similar in shape and normalization to the average scaled emission measure profile of nearby clusters.

In the following, we compare the emission measure profile of the main cluster of \object{RX~J0256.5+0006} with the emission measure profile of Arnaud, Aghanim \& Neumann (2002a) to see whether this cluster, dispite its distorted morphology follows the found self-similarity of clusters.

\subsubsection{From the surface brightness to the scaled emission measure profile}

The surface brightness profile is translated into the emission measure (EM) profile via~:
\begin{equation}
\mathrm{EM}(r)\propto (1+z)^{4}{{S(\theta)}\over{\epsilon(\mathrm{T},z)}}
\end{equation}

$\theta$ is the angle, $d_{\mathrm{A}}$ is the angular diameter distance, $r=\theta\times d_{\mathrm{A}}$, and $\epsilon(\mathrm{T},z)$ is the emissivity in the considered energy band (0.3 to 3.0\,keV in our case). The determination of $\epsilon$ takes into account the response of the cameras and galactic absorption(see also Sec.\ref{temp}).

The resulting emission measure profile is scaled to the self-similar model by using the standard scaling laws and the empirical $M_{gas}-T$ relation found by Neumann \& Arnaud (2001). The physical radius is scaled according to the calculated virial radius $r_{v}$ (see equ.\,\ref{evorv}-- please note: this value only depends on the measured cluster temperature and not on our determined mass profile) and the emission measure is scaled by~:
\begin{equation}
\mathrm{ScEM}\propto{\Delta_{z}}^{-{{3}\over{2}}}{(1+z)}^{-{{9}\over{2}}}T^{-1.38}
\end{equation}
where $\Delta_{z}$ is defined in equ.\,\ref{evorv}.

These scaling relations are applied to the three surface brightness profiles (see Sec.\ref{sbp}). The resulting scaled emission measure profiles are shown in Fig.\ref{emscp}.

\subsubsection{Comparison with the local reference profile}

\label{sec:scale}

\begin{figure}
\resizebox{\hsize}{!}{\includegraphics{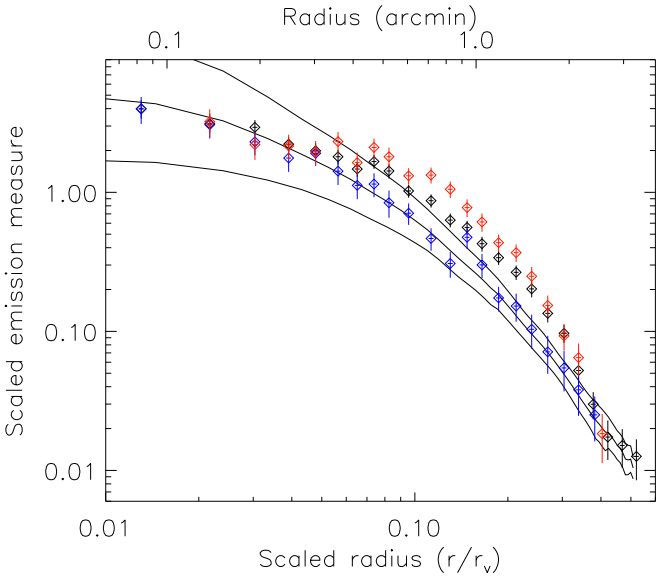}}
\caption{Comparison of the different scaled emission measure profiles of \object{RXJ0256.5+0006} with the reference profile (lines) based on nearby clusters from Arnaud et al. (2002a). The blue (black, red) circles show the NE (overall, SE) profile.  The error bars on the scaled emission measure are 1\,$\sigma$  taking into account statistical and temperature uncertainties.}\label{emscp}
\end{figure}

We show in Fig.\ref{emscp} the scaled emission measure profiles of the different cluster regions on top of the reference profile of nearby clusters (see Neumann \& Arnaud 1999). Only the NE profile (see Tab.\ref{massbeta}) matches the profile from the nearby clusters. The overall as well as the SE profile lie systematically too high with respect to the nearby cluster profile up to a radius of $0.35\times r_v$. This implies that only the NE profile shows self-similarity with other cluster profiles and at the same time that only the NE profile represents the relaxed part of the cluster. The two other profiles comprise (a) region(s), which are very likely disturbed due to past merging activity. The difference of the profiles between each other and with respect to the mean nearby cluster profile shows how important it is for self similar studies to select relaxed (parts of) clusters. 

\subsection{Total mass estimates using hydrostatic equilibrium}\label{totmass}

In the following we estimate the total mass of the cluster using the hydrostatic equilibrium hypothesis.

\begin{table*}
\begin{center}
\begin{tabular}{ccccccc}
\hline
\hline
\\
 & \multicolumn{6}{c}{Radial profiles}
\\
 & \multicolumn{2}{c}{Overall} & \multicolumn{2}{c}{NE} & \multicolumn{2}{c}{SE}
\\
 & $a$ & $b$ & $a$ & $b$ & $a$ & $b$
\\
\hline
$r_{500}$ (kpc) & 1410$^{+80}_{-70}$ & 1510$^{+100}_{-90}$ & 1210$^{+80}_{-70}$ & 1250$^{+250}_{-90}$ & 1760$^{+300}_{-200}$ & 1780$^{+350}_{-220}$
\\
$M_{500}$ (10$^{14}$\,M$_{\odot}$) & 5.9$^{+1.1}_{-0.9}$ & 7.3$^{+1.6}_{-1.2}$ & 3.7$^{+0.8}_{-0.6}$ & 4.1$^{+2.9}_{-0.8}$ & 11.5$^{+6.9}_{-3.4}$ & 11.9$^{+8.5}_{-3.9}$
\\
$M_{\mathrm{gas},500}$ (10$^{14}$\,M$_{\odot}$) & 1.10$\pm0.04$ & 1.18$\pm0.04$ & 0.73$\pm0.07$ & 0.76$^{+0.15}_{-0.21}$ & 1.44$\pm0.14$ & 1.43$\pm0.23$
\\
$f_{\mathrm{gas},500}$ (\%) & 18.6$^{+4.5}_{-3.5}$ & 16.2$^{+4.1}_{-3.2}$ & 19.7$^{+6.2}_{-5.1}$ & 18.5$^{+16.8}_{-8.7}$ & 12.5$^{+8.7}_{-4.9}$ & 12.0$^{+10.5}_{-5.8}$
\\
\hline
\end{tabular}
\end{center}
\caption{Total mass, gas mass and gas mass fraction at $r_{500}$ for the main cluster of \object{RX~J0256.5+0006}  based on hydrostatic equilibrium assuming isothermality. The quantities are calculated using the estimated mean temperature (Sec.\ref{temp}) and the different $\beta$ model best fit parameters (see Sec.\ref{sbp} and Tab.\ref{betasbp}). The uncertainties (90\% confidence level) take into account the errors on temperature and \betam parameters.}\label{massbeta}
\end{table*}

In the case of hydrostatic equilibrium, the gravitational mass $\mathrm{M}_{\mathrm{tot}}$ of a galaxy cluster can be written as~:
\begin{equation}
M_{\mathrm{tot}}(<r)=-{{kT}\over{G \mu m_p}}r\Bigg({{\mbox{d}\ln n_e}\over{\mbox{d} \ln r}}+{{\mbox{d} \ln T}\over{\mbox{d} \ln r}}\Bigg)\label{totalmassequ}
\end{equation}
where $k$ is the Boltzmann constant, $T$ the gas temperature, $G$ the gravitational constant, $\mu$ the mean molecular weight of the gas ($\mu\sim 0.6$), $m_{p}$ the proton mass and $n_e$ the electron density.

In Sec.\ref{hrmap} and \ref{temp}, we found no indication for important temperature variations, therefore we neglect the temperature gradient in equ.\ref{totalmassequ}. For our calculation of the mass of the cluster we use in the following temperature $kT=4.9^{+0.5}_{-0.4}$\,keV.

Including the \betam in equ.\ref{totalmassequ} we obtain:

\begin{equation}
M_{tot}(<r)={{3k \beta}\over{G \mu m_p}}T{{r^{3}}\over{r^{2}+r_c}^2}\label{mtotbeta}
\end{equation}

Since we have different fit results for the \betam parameters (depending on the region in which the surface brightness profile was extracted -- see also Sec.\ref{sbp}) we calculate in the following the mass of the cluster using the different model parameters. In order to be able to compare our results with other studies we calculate for the different $\beta$ models the mass in a region in which the mean density of the cluster has an over-density of 500 with respect to the critical density of the universe at the cluster redshift $\rho_{c}(z)$. The corresponding radius and mass for over-density 500 are defined as $r_{500}$ and $M_{500}$.  In a  $\Lambda$CDM Universe we can write for $\rho_{c}(z)$:

\begin{equation}
\rho_c(z) = \rho_{c0}(\Omega_{m}(1+z)^{3}+\Omega_{\Lambda})
\end{equation}
where $\rho_{c0}$ is the local value of the critical density (Weinberg 1972) and~:
\begin{equation}
\rho_{c0}={{3\,H^{2}_{0}}\over{8\pi G}}\sim 69.4\,M_{\odot}/\mathrm{kpc}^{3}
\end{equation}
At a redshift of 0.36, for $\Omega_{m}$=0.3 and $\Omega_{\Lambda}$=0.7, we find $\rho_{c}=100.9 M_{\odot}$/kpc$^{3}$.

Using equ.(\ref{mtotbeta}) at $r_{500}$ we can write~:
\begin{eqnarray}
M_{tot}(< r_{500})=M_{500} & = & {{3k\beta}\over{G\mu m_{p}}}T{r^{3}_{500}}\over{r^{2}_{500}+r_{c}^2}\nonumber
\\
 & = & {{4}\over{3}}\pi r^{3}_{500}\times 500\rho_c\label{mtotdensity}
\end{eqnarray}

We derive $r_{500}$ using equ.(\ref{mtotdensity}) for the different \betam parameters (see Tab.\ref{betasbp}). To keep a confidence level of 90\%, we propagate the corresponding errors from the cluster temperature and the $\beta$ model parameters.

\subsection{Comparison of total mass with other work}

We compare our result with the $M-T$ study presented by Finoguenov, Reiprich \& B\"{o}hringer (2001) based on the {\sc hiflugcs} galaxy cluster sample (Reiprich \& B\"ohringer 2002). Adding redshift evolution effects and transposing their found relation in our cosmological model we can write (neglecting uncertainties):

\begin{equation}
M_{500}=\frac{\Delta_z^{-1/2} (1+z)^{-3/2}}{\Delta_{z=0}^{-1/2}} 4.3 10^{13}\,M_{\odot} \left(\frac{kT}{\mbox{keV}}\right)^{1.58}\label{mt}
\end{equation}

and

\begin{equation}
\frac{r_{500}}{\mbox{Mpc}} = \frac{\Delta_z^{-1/2} (1+z)^{-3/2}}{\Delta_{z=0}^{-1/2}} 0.76\sqrt{\frac{kT}{\mbox{keV}}}
\end{equation}

Inserting our measured cluster temperature $kT=4.9$~keV, we find $M_{500}=3.9\times 10^{14}$ \msun and $r_{500}=1.2$~Mpc. These values are in good agreement with the values we found with the hydrostatic equilibrium (Tab.\ref{massbeta}). Using the NE profile, which very likely represents the relaxed part of the cluster (see Sec.\ref{sec:scale}), we found $3.7^{+0.8}_{-0.6}\times 10^{14}$\msun when we use the fit parameters taking into account the central part of the surface brightness profile and $4.1^{+2.9}_{-0.8}\times 10^{14}$\msun when neglecting the inner parts of the surface brightness profile.
Discarding evolutionary effects (neglecting the $(1+z)^{-3/2}$ term), we find for the above relations: $r_{500}=2.0$~Mpc and $M_{500}=6.2\times 10^{14}$~\msun. Not surprisingly, the calculated non-evolution $M_{500}$ and $r_{500}$ are much larger than the values we found using the best fit \betam parameters of the NE profile (Tab.\ref{massbeta}). The value for the no-evolution $r_{500}$ exceeds the calculated $r_{500}$ value even when taking into account the errors. The no-evolution $M_{500}$ agrees within the error bars only with the measured $M_{500}$ when the central bins are cut out. In this case the the calculated uncertainties are substantial (see Tab.\ref{massbeta}). We take this as an indication that $M_{500}$ and $r_{500}$ do in fact evolve with redshift.

\subsection{Gas mass and gas mass fraction}

We obtain the gas mass by integrating equ.\,(\ref{ne}) and using the $\beta$ model parameters determined in Sec.\ref{sbp}. The fraction of X-ray emitting gas of the main cluster of \object{RX~J0256.5+0006} is the ratio of gas mass to total mass,  which we determined in Sec.\ref{totmass}. Our results on the gas mass fraction are displayed in Tab.\ref{massbeta}.

The calculated gas masses depend on the determined $\beta$ model parameters. Since we find different $\beta$'s, not surprisingly, the different gas masses differ by a factor of 2. However, since the errors are relatively large, the gas mass fractions agree within the uncertainties. Our found values for $f_g$ agree within the error bars

with the value of 20.0$\pm$1.9\,\% obtained for hot clusters from Arnaud \& Evrard (1999). For the NE profile of the cluster, which seems to represent the relaxed part of the cluster (see above and Sec.\ref{sec:scale}), we find a gas mass fraction roughly between 18\,\% and 20\,\%, which is slightly lower but in good agreement with the value found by Arnaud \& Evrard (1999). 

Since the gas mass fraction in galaxy clusters should remain constant with redshift, Pen (1997) suggested that $f_g$ can be used to determine cosmological parameters.
If we adopt $\Omega_m=1$ and $\Omega_\Lambda=0$ the gas mass fraction lowers by about 15\%. This results in a gas mass fraction which is lower than the value found by Arnaud \& Evrard (1999). However, since our error bars are large we do not estimate this result significant enough to give any statement on the value of cosmological parameters.

\section{Optical observations of \object{RX~J0256.5+0006}}\label{optical}

\begin{figure}
\vspace*{5cm}
\caption{Optical image of the \object{RX~J0256.5+0006} sky region obtained at the 1.5\,m Danish telescope at ESO. The black vertical line is a dead column. The contours are the same as in Fig.\ref{signif}. The galaxies with spectroscopic redshift are labeled and each of them is tagged by a red circle. The redshifts are presented in Tab.\ref{redshifts}.}\label{optimg}
\end{figure}

The {\sc sharc} survey collected serendipitous extended sources from {\sc rosat} data. An optical follow-up campaign of the galaxy clusters was then undertaken (Romer et al. 2000, Burke et al. 2003). We present in this section the optical data obtained for \object{RX~J0256.5+0006}. The image which was obtained with the 1.5\,m Danish telescope at ESO is shown in Fig.\ref{optimg}. Redshifts of four galaxies (these galaxies present only absorption features, no emission lines were present in the optical spectra) were obtained with the 4\,m Kitt Peak telescope (see Tab.\ref{redshifts} and Fig.\ref{optimg}).

\begin{table}
\begin{center}
\begin{tabular}{cccc}
\hline
\hline
Label & Right ascension & Declination & Redshift
\\
\hline
A & 2$^{\mathrm{h}}$\,56$^{\mathrm{m}}$\,35.5$^{\mathrm{s}}$ & 00\deg\,06'\,10.8'' & 0.36148$\pm$0.00051
\\
B & 2$^{\mathrm{h}}$\,56$^{\mathrm{m}}$\,33.9$^{\mathrm{s}}$ & 00\deg\,06'\,40.9'' & 0.35774$\pm$0.00099
\\
C & 2$^{\mathrm{h}}$\,56$^{\mathrm{m}}$\,32.8$^{\mathrm{s}}$ & 00\deg\,06'\,25.3'' & 0.36451$\pm$0.00086
\\
D & 2$^{\mathrm{h}}$\,56$^{\mathrm{m}}$\,30.8$^{\mathrm{s}}$ & 00\deg\,06'\,03.5'' & 0.37044$\pm$0.00076
\\
\hline
\end{tabular}
\end{center}
\caption{Spectroscopic redshifts for the labeled galaxies (see Fig.\ref{optimg}) obtained at the 4\,m Kitt Peak telescope with a 1\,$\sigma$ confidence level (internal errors).}\label{redshifts}
\end{table}

The error bars obtained for the galaxy redshifts in Tab.\ref{redshifts} only account for the internal errors due to cross correlation analysis. The redshifts of galaxies B and D obtained from the {\sc sdss} (Sloan Digital Sky Survey) are 0.358 and 0.371, respectively. These values are in good agreement with the observed redshifts in Tab.\ref{redshifts} and indicate that the {\sc sharc} redshifts are correct.

\subsection{Galaxy position and X-ray emission correlations}\label{galpos}

Fig.\ref{optimg} displays many bright galaxies inside the region of the main X-ray cluster.

One galaxy, located approximately 8'' south of galaxy B coincides nicely with the central residual found in X-rays (see Sec.\ref{morph}). Since the extent of this residual is very small, it is not unlikely that the corresponding source is point-like, such as an {\sc agn} for example.

Galaxy D coincides well with the X-ray maximum of the subcluster. This galaxy is by far the brightest in this region. There is an offset of roughly 5'' between galaxy D and the maximum of the residuals. The projected offset corresponds to 40\,kpc. An explanation for this offset can be ram pressure stripping, which acts on the ICM but not on the galaxies during infall: unlike galaxies, which can be considered as collisionless matter particles which follow directly the dark matter distribution in cluster mergers, the hot gas of  the infalling subcluster is slowed down due to presence of the ICM of the main cluster (see numerical simulations performed by e.g. Roettiger, Loken \& Burns 1997; Stevens Acreman \& Ponman 1999). This effect was for example studied in Neumann et al. (2001) in the case of the Coma cluster. We will address the issue of ram pressure stripping in detail in Sec.\ref{sec:offset}.

\subsection{Velocity differences between main and subcluster}\label{velopt}

In the following we will estimate the velocity difference along the line-of-sight $v_{\bot}$ between the main cluster and the subcluster component based on the redshift measurements of the four galaxies (see also Tab.\ref{redshifts}).
xSince we only have the redshift for one galaxy in the subcluster, galaxy D ($z=0.3704$) we assume that this redshift is representative for the entire subcluster. 

Calculating the mean redshift of the other three galaxies (A, B, and C), assuming that they all belong to the main cluster and neglecting peculiar motions, we find $z=0.3612$. The difference in redshift between this mean redshift and the redshift of galaxy D corresponds to a velocity of 2750~km/s. In order to estimate the minimal $v_\bot$ we compare the redshift of galaxy D  with the redshift of galaxy C ($z=0.3645$), which has of the three galaxies A, B and C the redshift closest to galaxy D. The velocity difference corresponds in this case to 1750~km/s and is thus the minimal $v_\bot$. 

Since there exists very likely a velocity component parallel to the plane-of-sight, $v_\bot$ is only a lower limit on the actual infall velocity. In the following we will estimate the infall velocity and calculate a possible merger model for this cluster.  

\section{Dynamics of the merger in \object{RX~J0256.5+0006}}\label{dyncluster}

\subsection{Geometry of the merging cluster components}

\begin{figure}
\resizebox{\hsize}{!}{\includegraphics{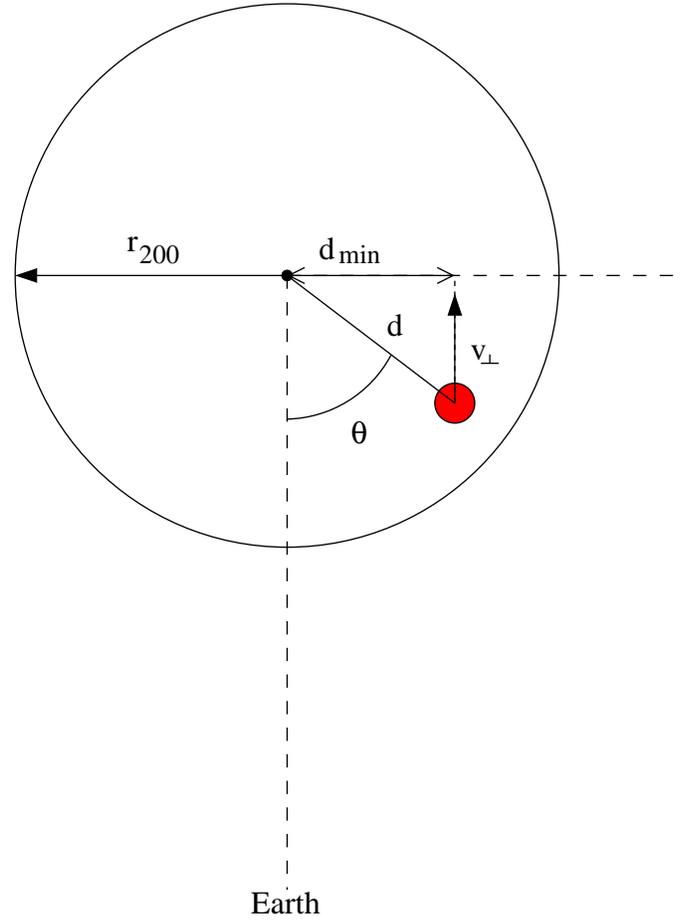}}
\caption{Merger geometry of \object{RX~J0256.5+0006} in the case of an on-axis merger. The small point in the centre represents the main cluster centre, and the larger point represents the subcluster. $d_min$ is the projected distance between the two cluster centers (see Sec.\ref{morph}), $d$ is the physical distance between the two components) and $v_{\bot}$, the projection of the impact velocity along the line-of-sight (see Sec.\ref{velopt}). We estimate that the main cluster extends up to $r_{200}$. $\theta$ is the impact angle.}\label{geometry}
\end{figure}

In the following we will try to quantify the merger geometry in this cluster by using the estimates of the projected distance between the two clusters $d_{min}$ (350\,kpc --- see also Sec.\ref{morph}) and v$_{\bot}$ (Sec.\ref{velopt}). The adopted merger geometry is shown in Fig.\ref{geometry}.

A similar approach has been already performed by e.g. Beers, Geller \& Huchra (1982) for Abell 98 and Colless \& Dunn (1996) for the Coma cluster.

\subsubsection{The merger model}\label{model}

To determine the merger geometry analytically, we apply the following assumptions~:
\begin{description}
\item[-] a purely gravitational two-body interaction, i.e. pressure free and no neighborhood effects~;
\item[-] the main component is defined by the mass distribution $M_tot(d)$ (see equ.\,(\ref{mtotbeta}))~;
\item[-] the subcluster mass distribution can be approximated with a point~;
\item[-] The main cluster is at rest, only the subcluster moves.
\end{description}

From the law of gravitation follows~:
\begin{equation}
{d\vec{v}\over{dt}} = -{GM_{tot}(d)\over{d^{2}}}\vec{i}\label{rfd}
\end{equation}
$\vec{v}$ is the velocity of the subcluster, $d$ is the distance between main cluster center and subcluster. $\vec{i}$ is an unity vector along the merger axis and is oriented in the direction of increasing $d$.

Using for simplicity that the distance of the two cluster components was close to infinity at the beginning we can distinguish between two scenario~:
\begin{description}
\item[-]$d>r_{200}$~: (numerical simulations suggest that $r_{200}$ corresponds to the virialized part of a cluster) in this case, the main cluster can be approximated with a point mass located at the cluster center. The velocity is~:
\begin{equation}
v = \sqrt{{{2G M_{200}}\over{d}}}
\end{equation}
\item[-]$d<r_{200}$~: in this case, we have to take into account the mass distribution defined in equ.\,(\ref{mtotbeta}). By integrating equ.\,(\ref{rfd}), we can write~:
\begin{equation}
v = \sqrt{{{2 G M_{200}}\over{r_{200}}}-{{3k\beta T}\over{\mu m_{p}}}\ln\Bigg({{d^{2}+r_c^{2}}\over{r_{200}^{2}+r_c^{2}}}\Bigg)}
\label{equ:v}
\end{equation}
\end{description}

\begin{figure}
\resizebox{\hsize}{!}{\includegraphics{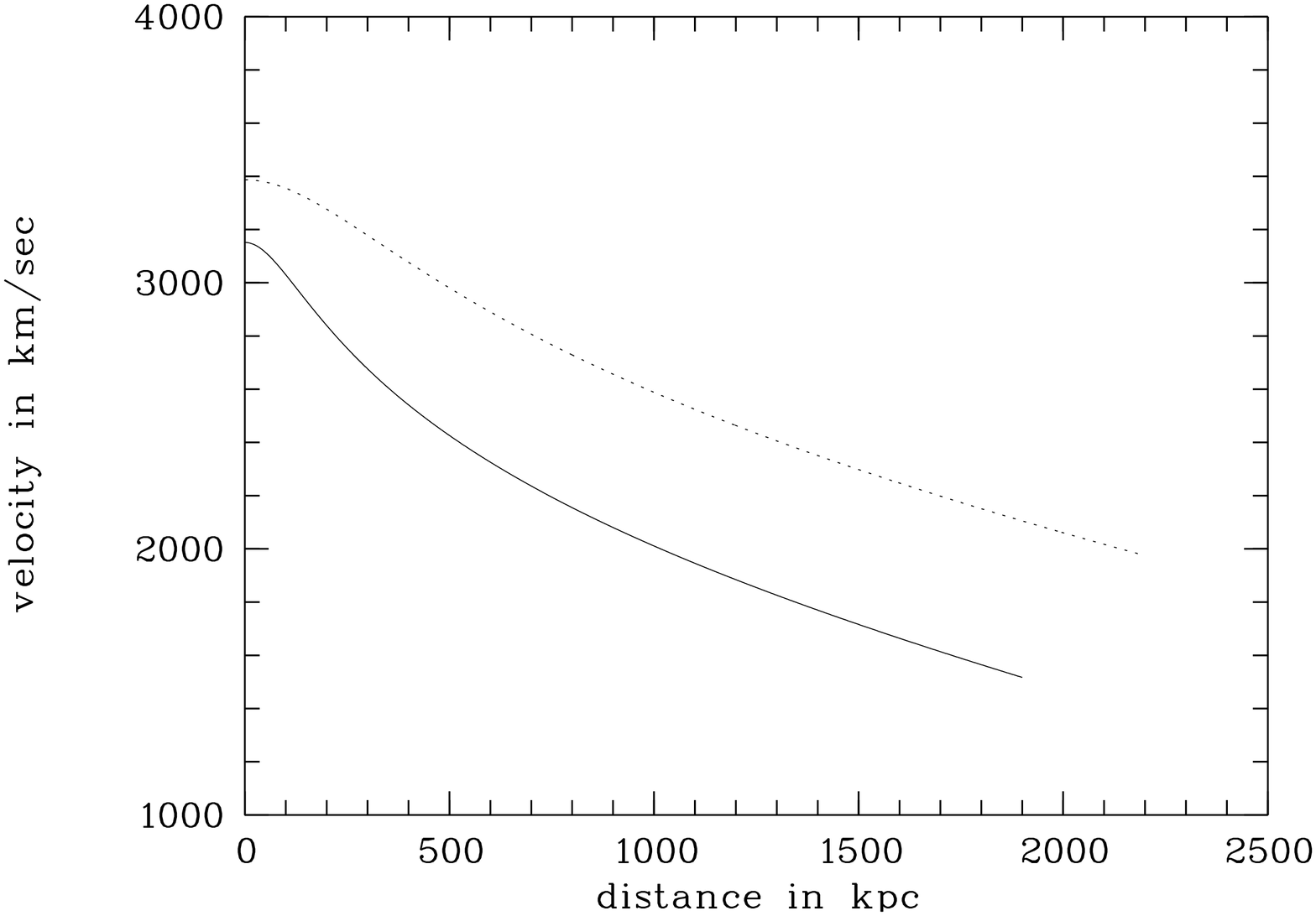}}
\caption{The velocity of the infalling subcluster as function of its distance $d$ to the main cluster centre. The full line represents the velocity when using the mass profile of the \betam parameters from the NE surface brightness model (see also Tab.\ref{betasbp}). The dotted line shows the velocity when using for the mass distribution of the main cluster the \betam parameters from the overall surface brightness profile  with the subcluster cut out.}\label{vr}
\end{figure}


To determine $r_{200}$ we apply equ.\,(\ref{mtotdensity}) with an over-density of 200. For the main cluster mass distribution as well as $r_{200}$ and $M_{200}$, we use the results of the \betam fits. More specifically we use in the following the fit results of the NE surface brightness profile and the overall profile taking into account the central part of the cluster (see Tab.\ref{betasbp} -- the $a$ case). The comparison of the results based on these two models gives an idea on the dependence of the merger dynamics on the mass profile. We give more weight on the results of the NE profile, since it seems to represent the relaxed part of the main cluster in which the hypothesis of hydrostatic equilibrium is most likely valid. 

 Using the overall model without angular selection we find  $M_{200}=9.2\times$10$^{14}$\,$M_{\odot}$ and $r_{200}=2.2$\,Mpc. For the NE model we find $M_{200}=5.9\times$10$^{14}$\,$M_{\odot}$ and $r_{200}=1.9$\,Mpc.

Fig.\ref{vr} shows the obtained relationship between distance $d$ and velocity $v$ for the two models.

\subsubsection{Impact angle and physical distance}\label{angle}

From Fig.\ref{geometry} follows~:
\begin{eqnarray}
\sin\theta & = & {{d_{min}}\over{d}} \label{d1}\\
\cos\theta & = & {{v_{\bot}}\over{v}}
\end{eqnarray}
$\theta$ is the impact angle, which is defined as the angle between the line-of-sight and the merger axis. The merger axis is the line which connects the two cluster centers.

Since we observe a comet-like structure for the subcluster (which indicates gas stripping, see also below), we assume that the two cluster components are interacting which implies $d<r_{200}$.

The redshift of galaxy D (see Tab.\ref{redshifts}) which we assume to be located roughly at the center of the subcluster is greater than the redshifts obtained inside the main cluster. This implies that the subcluster is in front of the main component and thus that $\theta$ must be lower than 90\deg (see Fig.\ref{geometry}).

\begin{figure}
\resizebox{\hsize}{!}{\includegraphics{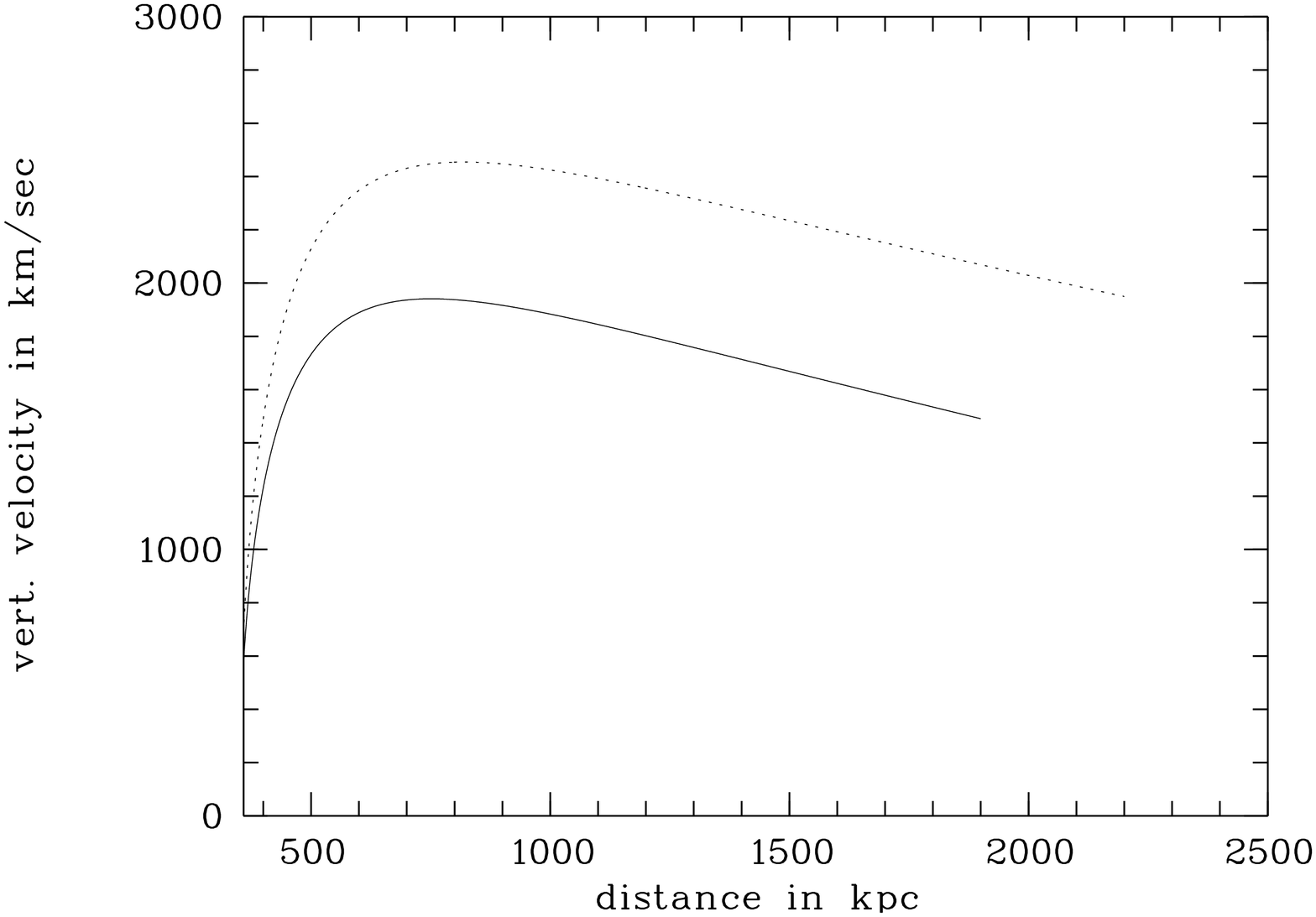}}
\caption{The vertical velocity component $v_{\bot}$ against the cluster distance according to the model described in Sec.\ref{model} and equ.\,(\ref{alphaequ}). Full and dotted line follow the same convention as Fig.\ref{vr}}\label{vperpd}
\end{figure}

The impact geometry can be furthermore restricted by combining equ.(18) and equ.(19) to:

\begin{equation}
{{d^{2}_{min}}\over{d^{2}}}+{{v^{2}_{\bot}}\over{v^{2}}}=1\label{alphaequ}
\end{equation}

The relationship between $v$ and $d$ is shown in Fig.\ref{vr}. Since we do not possess a precise estimate of $v_{\bot}$, we show in Fig.\ref{vperpd} the relation between $v_{\bot}$ and $d$. We can set the following limits: $d_{min}<d<r_{200}$ (Sec.\ref{morph}) and 1750~km/sec$<v_{\bot}<$2750~km/sec (Sec.\ref{velopt}). To constrain $v_{\bot}$ more precisely we would need more galaxy redshift measurements. 

\begin{figure}
\resizebox{\hsize}{!}{\includegraphics{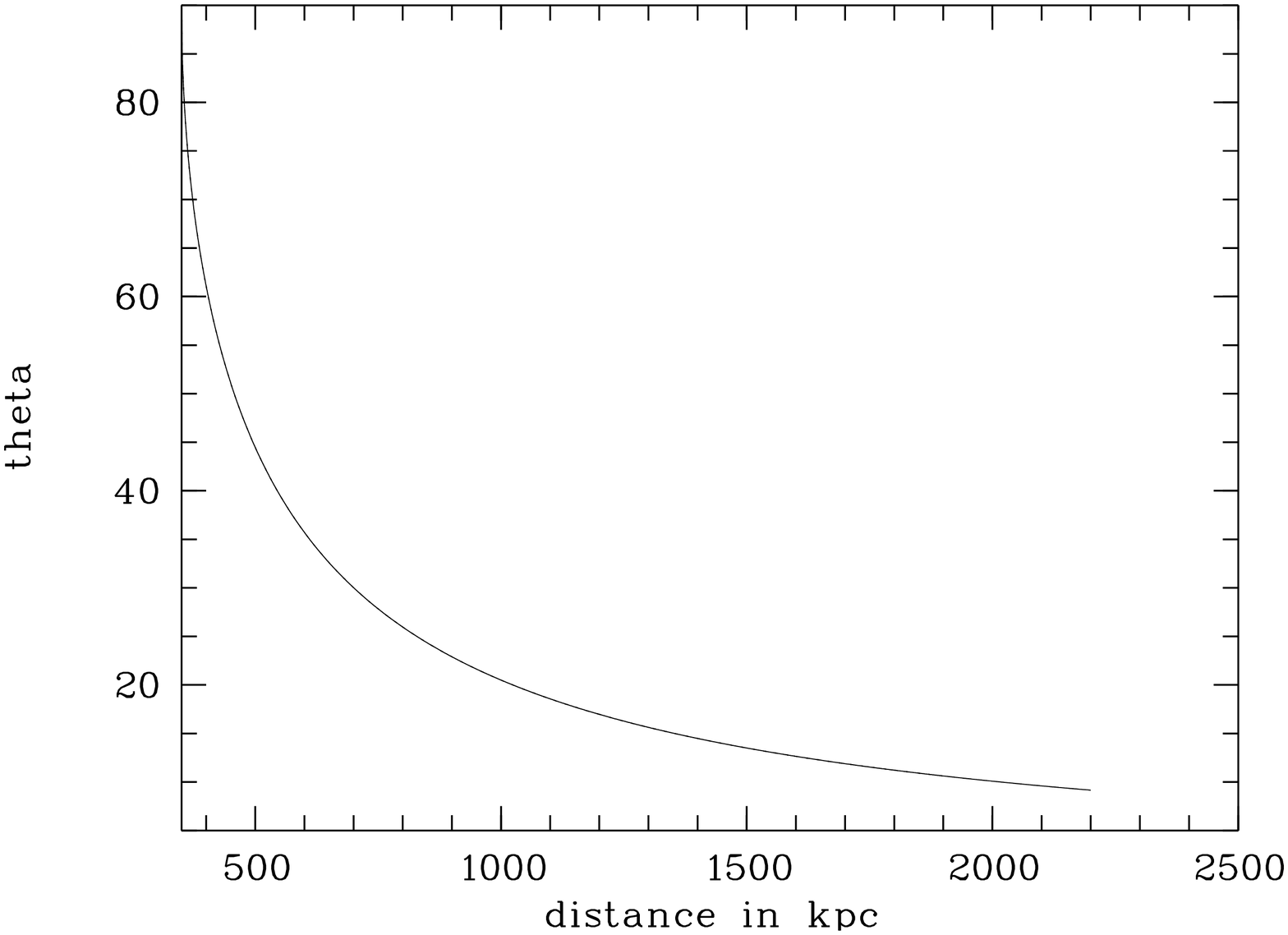}}
\caption{The impact angle $\theta$ defined in Fig.\ref{geometry} as function of distance $d$ between the two merger components. Since the projected distance is 350~kpc, $d\geq 350$~kpc. -- Please note: the function is independent on the cluster mass.}\label{ad}
\end{figure}

\begin{figure}
\resizebox{\hsize}{!}{\includegraphics{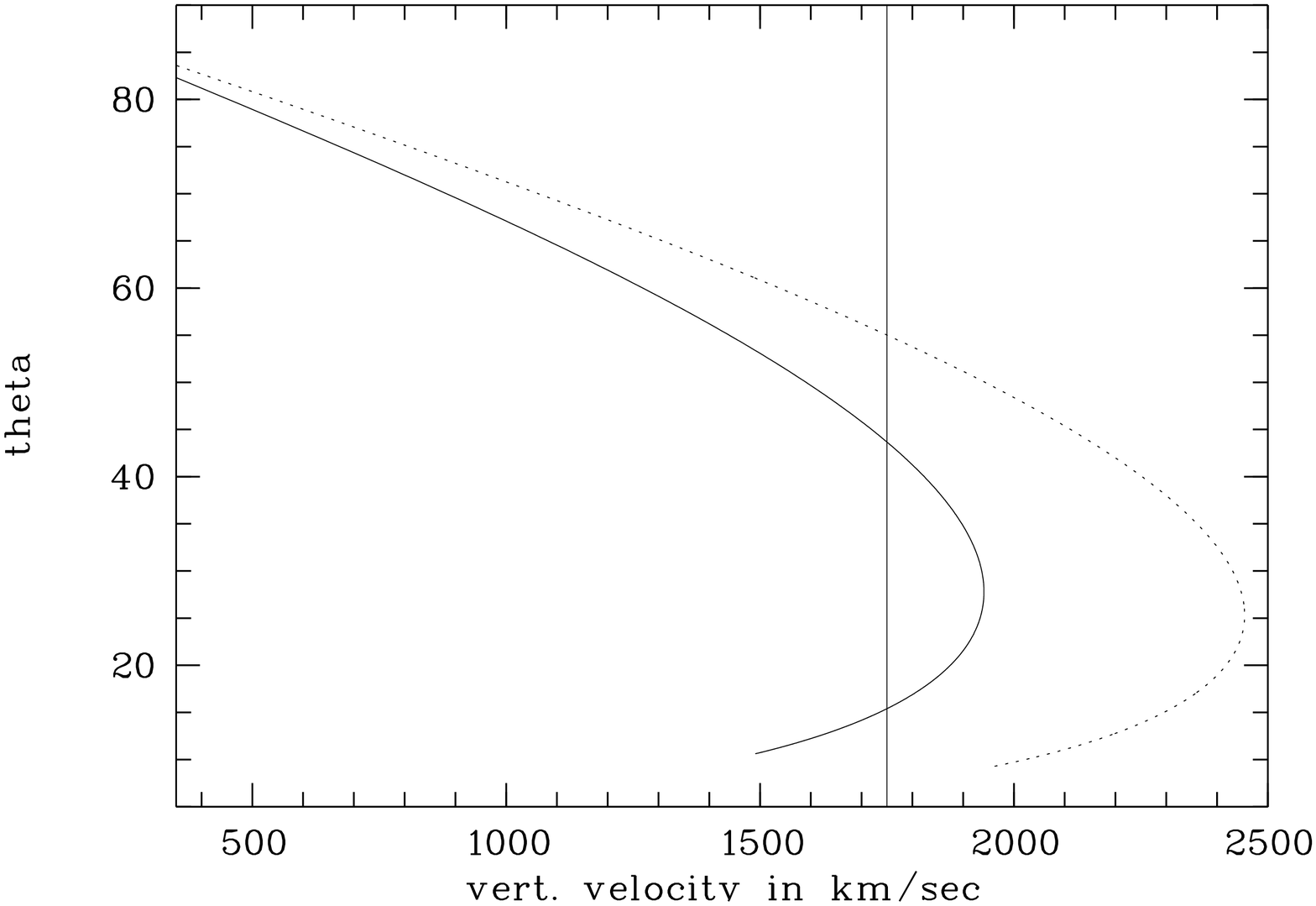}}
\caption{Impact angle $\theta$ defined in Fig.\ref{geometry} versus  $v_{\bot}$. Full and dotted line follow the same convention as Fig.12. The vertical line shows the minimal value for $v_\bot$.}\label{avperp}
\end{figure}

Fig.\ref{ad} and Fig.\ref{avperp} show the dependence of the impact angle $\theta$ on cluster distance $d$ and $v_{\bot}$. The greater $\theta$ is, the smaller is $d$. Based on our constraints on $v_{\bot}$ (vertical line in Fig.\ref{avperp}), we obtain values for $\theta$ between 15\deg and 44\deg (9\deg and 60\deg) for the NE (overall) mass model. Fig.\ref{avperp} shows that for each of the two adopted mass models any given value of $v_{\bot}$, gives two possible solutions for $\theta$. This means that a measured value of $v_{\bot}$ gives two different values of $\theta$ connected to two possible physical distances $d$. For $15^\circ < \theta < 44^\circ$ we find for the distance $d$: 0.5~Mpc$<d<$1.4~Mpc for the NE model. 

\subsubsection{Merger model dependence on main cluster mass distribution}

We see that the smaller the total mass of the main cluster is, the tighter are the constraints on the geometry of our merger model and angle $\theta$. This is not surprising since the total infall velocity is a function of cluster mass. Smaller masses show generally smaller infall velocities. For the NE mass model the observed vertical velocity $v_\bot$ is close to the maximal possible $v_\bot$ of about 1900~km/s. For comparison, for the overall mass model case the maximum velocity is roughly 2500~km/sec.

\subsection{The cometary shape of the subcluster: effects of ram pressure stripping}

The subcluster has a clear cometary shape, which suggests that ram pressure due to the infall plays an important role. In the following we will estimate the importance of ram pressure stripping in the merger model.
Ram pressure can be defined in our case as: 

\begin{equation}
p_{rp}=\rho_g v^2
\end{equation}

\begin{figure}
\resizebox{\hsize}{!}{\includegraphics{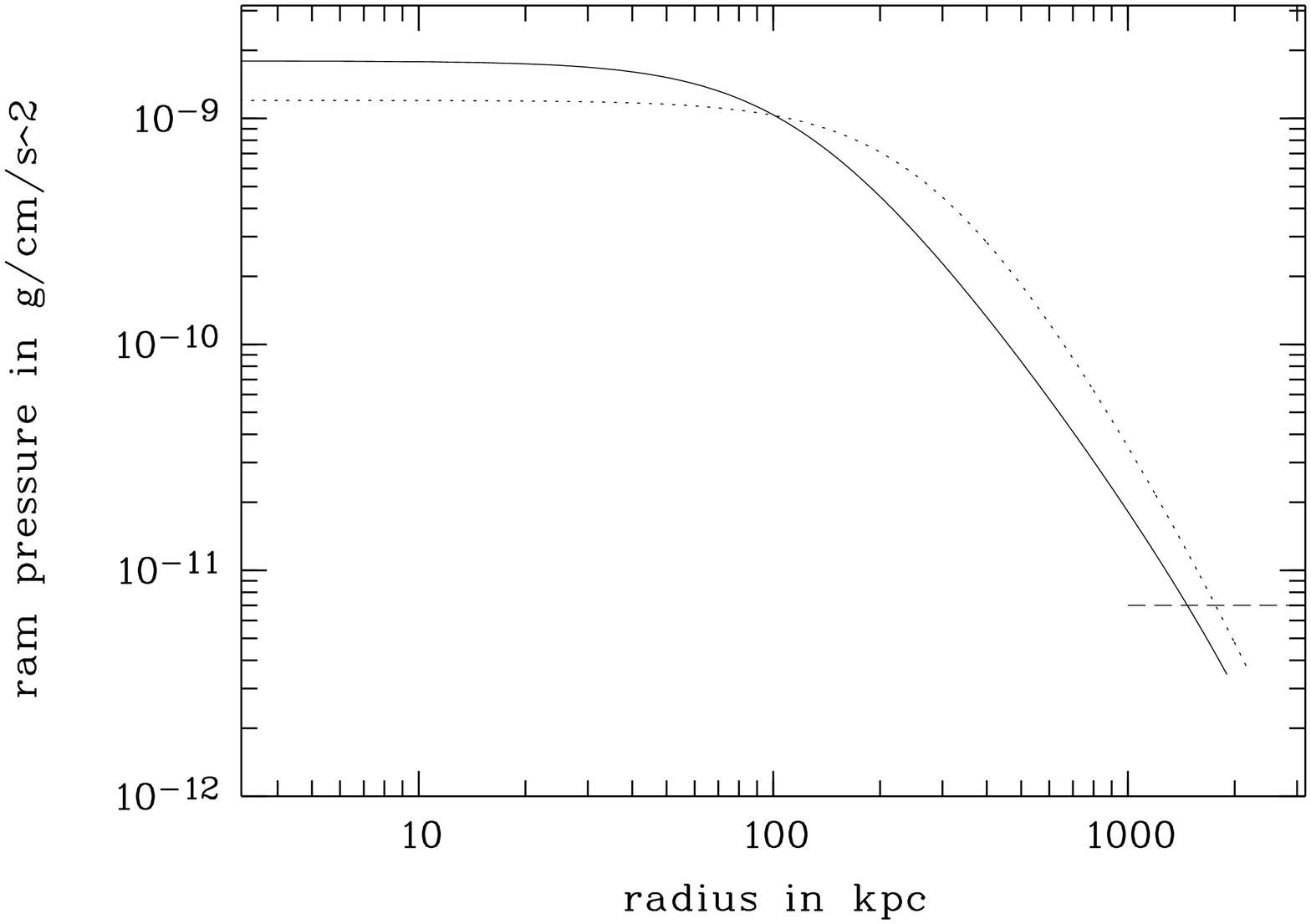}}
\caption{The ram pressure as a function of distance to the main cluster centre using equ.\ref{equ:v} and the gas density distribution based on the \betam parameters. The full line corresponds to the NE and  the dotted line to the overall \betam parameters. The horizontal dashed line indicates the internal pressure of the subcluster if we assume a constant electron gas density  of $n_e=7.1\times10^{-3}$cm$^{-3}$ (see also Sec.\ref{corlumin}).}\label{fig:ramp}
\end{figure}

$\rho_g$ is the gas density of the main cluster ($\rho_g=n \mu m_p; \mu =0.6; $ $m_p$: proton mass) and $v$ is the infall velocity of the subcluster with respect to the main cluster centre. The calculation of the merger geometry suggests that the subcluster has a physical distance of the main cluster of 0.5~Mpc$<d<$1.4~Mpc. Effects of ram pressure are important if it is of the same order or larger than the internal pressure of the subcluster. The internal pressure of the subcluster $p_{sg}=nkT=7\times 10^{-12}\mbox{g cm}^{-1}\mbox{s}^{-2}$ assuming an electron number density of $n_e(\mbox{subcluster})= 7.1\times 10^{-4}\mbox{cm}^{-3}$ (see Sec.\ref{corlumin}) and an internal temperature of the subcluster of about 3~keV (3~keV is the estimated temperature of the subcluster when using the z-evolving $L_X-T$-relation and the measured $L_X=1.9\times 10^{44}\mbox{erg/s}$ of the subcluster found below). Using as a typical infall velocity of $v=2000\mbox{km/s}$ we find that $\rho_g$ must be larger than $1.8\times 10^{-28}\mbox{g cm}^{-3}$ so that $p_{rp}>p_{sg}$. Using the NE (overall) $\beta$-model parameters we find that for example at 1~Mpc: $\rho_g=4.5\times10^{-28}\mbox{gcm}^{-3}$ ($\rho_g=5.5\times10^{-28}\mbox{gcm}^{-3}$). Since these values are larger than  $1.8\times 10^{-28}\mbox{gcm}^{-3}$ we conclude that ram pressure is actually important at this merger state. We display in Fig.\ref{fig:ramp} the ram pressure as function of distance to the main cluster centre for the overall and the NE profile. We use equ.(\ref{equ:v}) as velocity input for the calculation of the ram pressure in equ.\ref{equ:v} (see also Fig.\ref{vr}). We equally show the value of the internal pressure of the subcluster assuming a constant gas density distribution as horizontal line in Fig.\ref{fig:ramp} for comparison.

\begin{figure}
\resizebox{\hsize}{!}{\includegraphics{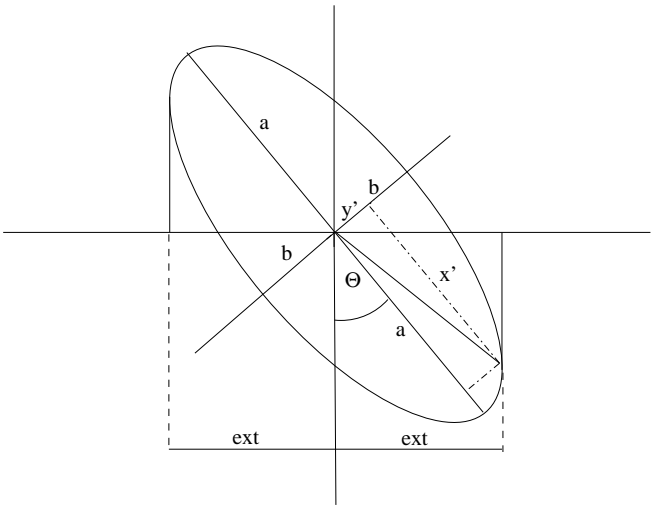}}
\caption{The ellipse model of the subcluster from above, looking down from North to South. 2x is the observed projected length of the subcluster.}
\label{fig:ell}
\end{figure}

In our adopted merger geometry, in which we find for the impact angle $15^\circ<\theta < 44^\circ$, the physical length of the subcluster is much larger than its projected length. If the subcluster for example had the form of a line then for $\theta = 20^\circ$, which corresponds to a physical distance of 1~Mpc the real length of the subcluster was three times larger than its projected length. The observed length is approximately 700~kpc, which implies a physical length in case of a line of approximately 2~Mpc (this is a factor of 3 larger than the width --600~kpc-- of the structure). However, it is very unlikely that the subcluster has the form of a line, it is more likely for it to be ellipsoidal. In this case the calculation of the physical length of the subcluster is more complicated. In the following we attempt to calculate the length of the subcluster assuming an ellipsoidal shape. Fig.\ref{fig:ell} shows our adopted model for the ellipsoid from above looking downwards (from North to South). 

We use for the calculation of the extent of the subcluster the ellipse equation:
\begin{equation}
\frac{x'^2}{a^2}+\frac{y'^2}{b^2}=1
\label{equ:elli}
\end{equation}

in which $x'$ and $y'$ are the x and y coordinates before rotation around the impact angle $\theta$. We furthermore use for the rotation around $\theta$:

\begin{equation}
x = x'\sin \theta + y' \cos \theta = x \sin \theta + \sqrt{1-\frac{x'^2}{a^2}}b\cos\theta
\label{equ:rot}
\end{equation}

Where $x$ is the x-coordinate after rotation. $ext$ (see Fig.\ref{fig:ell} is the maximum value in the x-direction after rotation. We can therefore write:
\begin{equation}
\frac{\mbox{d}(x = ext)}{\mbox{d}x'}=0
\label{equ:grad}
\end{equation}
Using equ.\ref{equ:rot} and equ.\ref{equ:grad} we find:

\begin{equation}
a = \frac{x'}{\sqrt{2}}\sqrt{1+\sqrt{1+4\frac{b^2}{ext^2\tan^2\theta}}}
\label{equ:a}
\end{equation}

Inserting equ.\ref{equ:a} in equ.\ref{equ:rot} we find

\begin{equation}
ext = x'\sin \theta + \sqrt{1-\frac{2}{1+\sqrt{1+\frac{4b^2}{x'^2\tan^2\theta}}}}
\label{equ:xxi}
\end{equation}

\begin{figure}
\resizebox{\hsize}{!}{\includegraphics{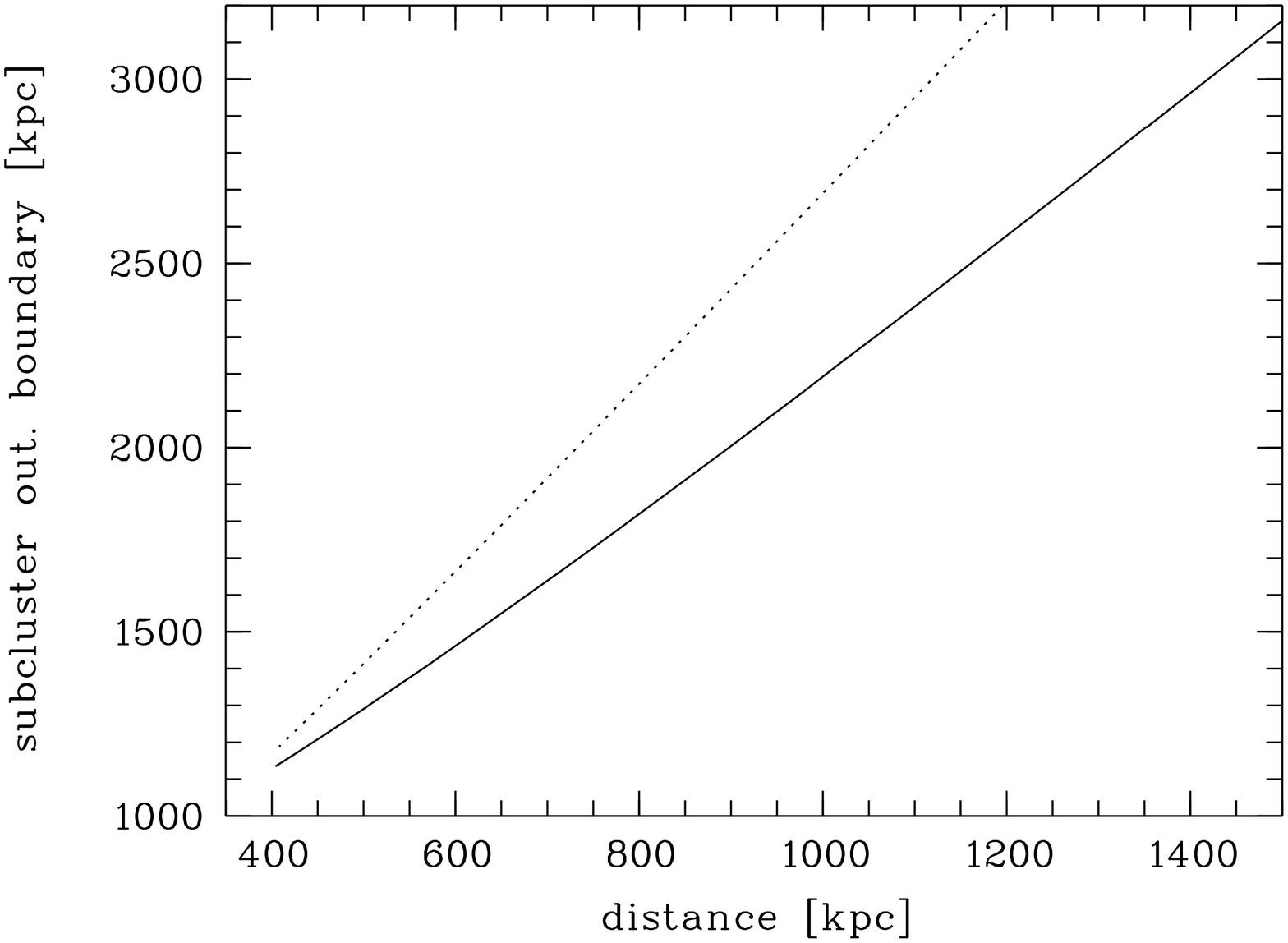}}
\caption{The subcluster outer boundary with respect to the main cluster centre  ($c=2a+d$) as function of subcluster main cluster distance $d$ (see also text).}\label{fig:ellext}
\end{figure}

Unfortunately this equation is too complicated to calculate $x'$ directly as function of $\theta$. However, knowing $ext$ and $b$ (we assume symmetry around the merger axis, so $b=300$~kpc and $ext=350$~kpc see also Sec.\ref{morph}) we can search for a given of $\theta$ the corresponding $x'$ with the requirement that $ext$=350~kpc. Subsequently we can calculate $a(x')$. $2a$ is the physical extent of the subcluster along the merger axis. 
$c=2a+d$ is the distance of the outer boundary of the subcluster to main cluster centre ($d$ and $\theta$ are correlated according to Fig.\ref{ad}). From Fig.\ref{fig:ramp} we can give limits on $c$. Applying the internal pressure argument of the subcluster from above we find a maximum value for $c$ of 1.5~Mpc. If we assume that $r_{200}$ defines the area in which ram pressure plays an important role then we find a maximum of $c=1.9$~Mpc. Fig.\ref{fig:ellext} shows $c$ as function of distance to the cluster centre. We find that distances up to 0.9~Mpc are possible in case of $r_{200}$ and 0.6~Mpc in the case of $c=1.5$~Mpc.  If we assume for example that $2b=400$~kpc instead of 600~kpc, we find the results indicated as dotted line in Fig.\ref{fig:ellext}. In this case we find smaller values for $d$. This shows the dependence of our calculations on the choice of $b$. The smaller $b$, the smaller is the allowed range of $d$. However, it seems rather unlikely that $2b=400$~kpc, which would imply a relatively large asymmetry of the subcluster. Ram pressure stripping makes the substructure more symmetric around the axis of infall (the parts of the subcluster with the largest distance with respect to its centre are less bound to the structure and are thus more easily stripped). Since ram pressure is efficient in this merger it is very likely that the symmetry around the infall axis is already established.

We conclude that due to observed effects of ram pressure stripping that the subcluster has a distance to the main cluster of roughly 0.6--0.9~Mpc.

\subsection{The offset between X-ray maximum and galaxy D}
\label{sec:offset}

We have seen that ram pressure stripping is efficient in the subcluster as it falls onto the main cluster. Ram pressure acts mostly on the gas and much less on galaxies. Therefore there should be a displacement between the subcluster ICM and galaxy D, which we assume traces the centre of gravity of the subcluster (which is suppossedly dominated by dark matter). We actually see a displacement of about $5''$ (see Fig.\ref{signif}). We can estimate the expected displacement between gas and galaxy D using 
\begin{equation}
M_{gas} v_{dec} =  \int M_{gas} dec dt = 
\int p_{rp} S \frac{dt}{dr}dr  = 
\\
\int\frac{ p_{rp} S}{v} dr 
\label{equ:decel}
\end{equation}
with $M_{gas}$: gas mass of subcluster; $v_{dec}$ the difference of velocity between galaxy D and gas; $dec$: deceleration of the gas with respect to the subcluster centre; $S$: Surface of subgroup facing the centre of the main cluster. We assume that the gas density is constant within the subcluster. We integrate equ.\ref{equ:decel} numerically. We use for $M_{gas}$ $10^{13}$\msun, which is about 10\% of the total mass found for the subcluster (see Sec.\ref{corlumin}). For $S$ we use a circle with radius 300~kpc(see above), which likely matches half of the diameter of the structure in width. $v_{dec}$ is inversely proportional to $M_{gas}$ and directly proportional to $S$. In order to estimate the displacement of the centre of the gas with respect to galaxy D, we integrate

\begin{equation}
d_{disp} = \int v_{dec} dt = \int v_{dec} \frac{dt}{dr} dr = \int \frac{v_{dec}}{v}dr
\end{equation} 

\begin{figure}
\resizebox{\hsize}{!}{\includegraphics{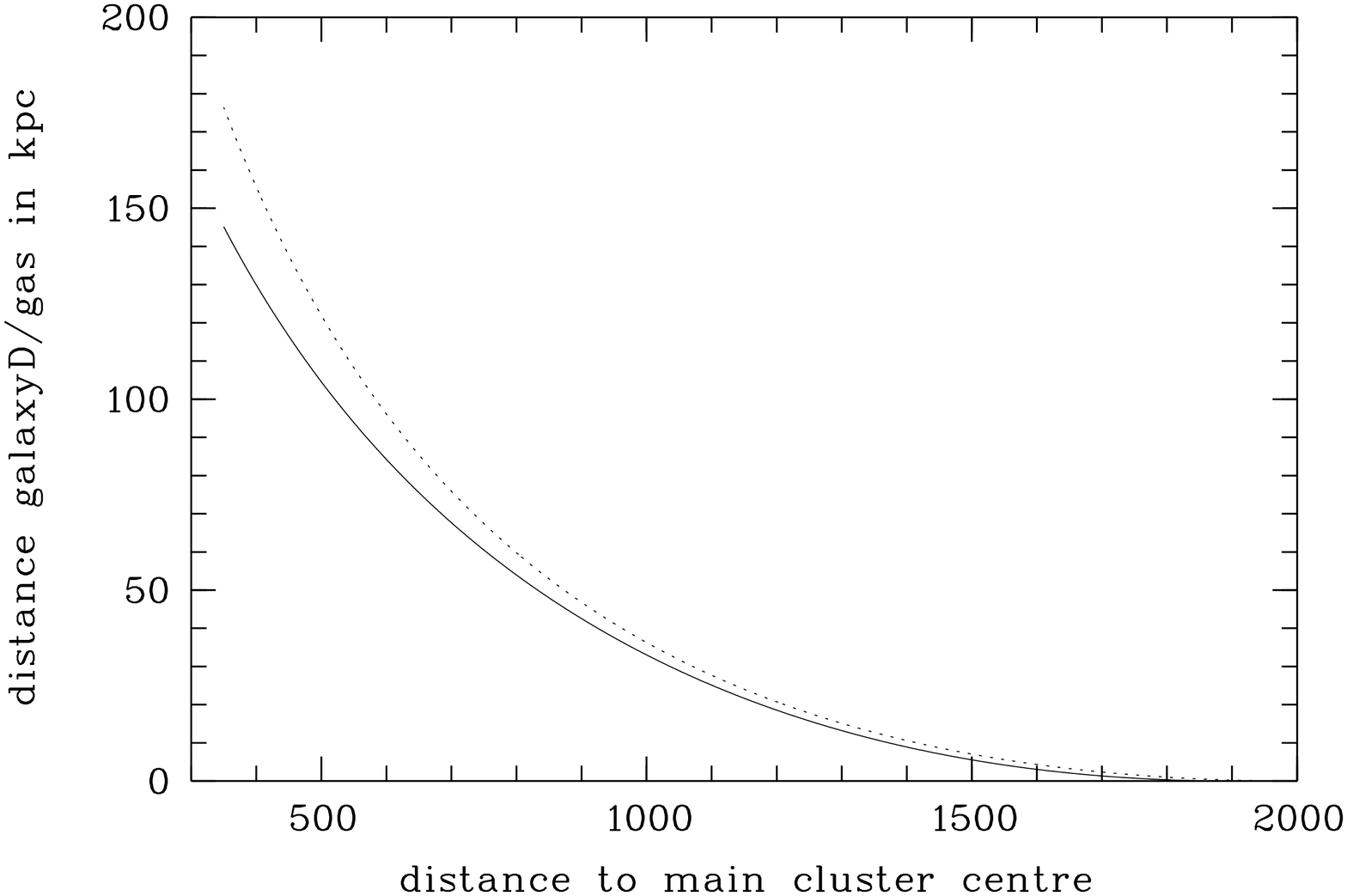}}
\caption{The physical distance between galaxy D and the maximum of X-ray emission of the subcluster based on calculations of ram pressure stripping as function of distance to the main cluster centre. Full line: calculations based on NE \betam; dotted line: based on overall \betam.}\label{dispkpc}
\end{figure}

\begin{figure}
\resizebox{\hsize}{!}{\includegraphics{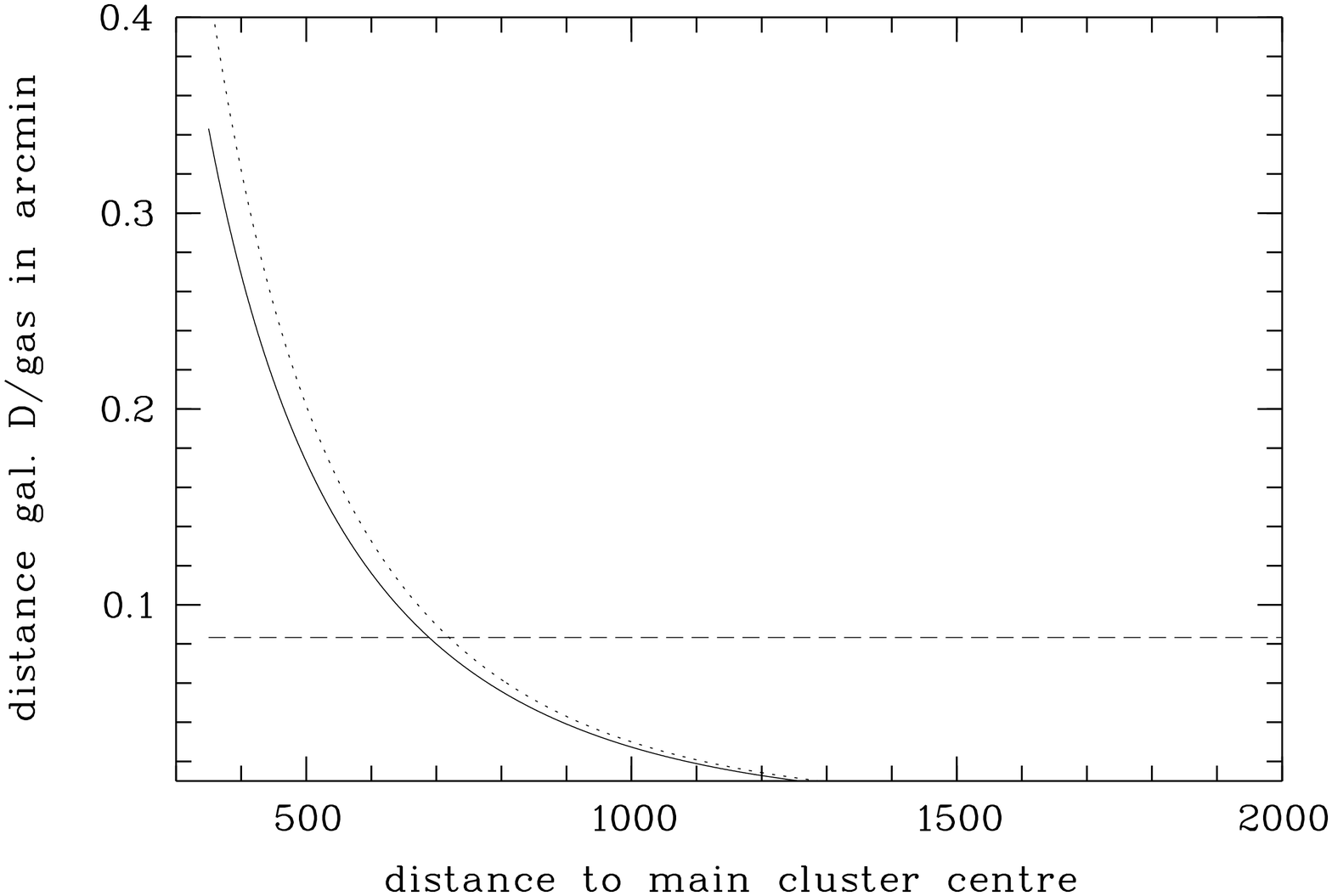}}
\caption{The projected distance between galaxy D and the maximum of X-ray emission of the subcluster based on calculations of ram pressure stripping as function of distance to the main cluster centre. Full line: based on NE \betam; dotted line: based on overall \betam. The horizontal dashed line shows the observed offset of galaxy D to subcluster X-ray maximum.}\label{dispamin}
\end{figure}

Again, we integrate numerically to calculate $d_{disp}$ the distance between galaxy D and the location of the X-ray residual. In Fig.\ref{dispkpc} we show $d_{disp}$ as a function of the distance of the subcluster to the main cluster centre. Fig.\ref{dispamin} shows the projected distance of galaxy D and subcluster in arcmin as function of main cluster distance. The full (dotted) line corresponds to the NE (overall) mass model. The horizontal dashed line corresponds to the displacement observed in Fig.\ref{optimg} of approximately 5'', which implies a distance of subcluster to main cluster centre of about 700~kpc. It is difficult to assess the correct offset between the X-ray maximum of the subcluster and galaxy D. Several parameters can in principle play a role: the parameters of the subtracted \betam and uncertainties of the alignment of X-ray image and optical image. We vary the parameters of the subtracted model within the error bars and see that the location of the X-ray residual maximum does not change. We estimate the alignment uncertainties, which are correlated with PSF effects in the order of half of the PSF size of \xmm, which is about 2-3~arcsec. Therefore, we estimate the maximum allowed offset between X-ray maximum and galaxy D to be 8~arcsec, or 0.133~arcmin. The implied minimal distance of the subcluster to the main cluster centre in our simple model is thus about 600~kpc (see also Fig.\ref{dispamin}). Inversely, the minimal offset between galaxy D and the subcluster X-ray emission is about 2~arcsec. Using Fig.\ref{dispamin}, this corresponds to a distance of the subcluster to the main cluster of about 1~Mpc. 

Therefore, the distance of the subcluster to the main cluster calculated from a simple ram pressure stripping model varies between 0.6 to 1~Mpc. This suggests that the impact angle $\theta$ lies between 20 to 35$^\circ$. This is in good agreement with our merger model. We would like to stress, however, that the model we adopted for ram pressure is very simple and is based on many assumptions. 

\subsection{Constraints on impact parameter}

We find coherent results for the distance of the subcluster to the main cluster centre using our on-axis merger model and calculating effects of ram pressure. With all this information we are in principle able to put some constraints on the probability that the merger is off-axis with a non-zero impact parameter. Most possible scenarios for an off-axis merger would increase the visible effects of ram pressure stripping since off-axis merger trajectories are generally longer than on-axis merger trajectories. Therefore there is more time for ram pressure to be efficient. We would thus expect that ram pressure calculations give distances of the subcluster to the main cluster, which are smaller than distances calculated with the on-axis merger model. However, this is not the case here. Remaining possible models are those in which  beside the infall velocity related to the gravitational force there exists an additional  velocity component in direction of the line-of-sight and perpendicular to it with a specific relationship between the velocity components. In this particular case, by chance, the merger would have a non-zero impact parameter with a specific orientation, which mimics at the moment of observation an on-axis merger with the correct projected elongation of the substructure caused by ram pressure. It is doubtful whether this is likely. A simple explanation of an on-axis merger seems to us much more realistic.

\subsection{Subcluster mass estimate}\label{corlumin}

Determining the luminosity of the subcluster is more difficult than for the main cluster (see Sec.\ref{lumin}) since we cannot fit an adequate model to the comet-like structure. To overcome this problem, we simply determine the count rate in the region of the residual map (after main cluster emission subtraction, see also Sec.\ref{morph}) in which the subcluster is clearly detected. Translating the count rate into luminosity, we find $L_{bol,{\sc x}}=1.9\pm0.3\times10^{44}$\,ergs/s.

Since we use the luminosity estimate via the scaling law relations to obtain the mass of the subcluster, luminosity biases introduce biases in our mass estimate. The luminosity estimate of the subcluster can be biased so that~:
\begin{description}
\item[-]we underestimate the total count rate since we only take into account the region in which we detected the subcluster. This subcluster might be extended to larger regions for which we do not account.
\item[-]since the bremsstrahlung emission evolves with $n_e^2$ and because the subcluster lies very likely inside the main cluster gas sphere, the main cluster emission model subtraction is not correct to obtain the subcluster emission. In the case that the subcluster is in the gas sphere of the main cluster, the total emission evolves with $n_e^2=(n_{e,1}+n_{e,2})^{2}>n_{e,1}^{2}+n_{e,2}^{2}$ where $n_{e,1}$, is the electron density of the main cluster, and $n_{e,2}$ the electron density of the subcluster. The quantity $2n_{e,1}n_{e,2}$ enhances the measured luminosity. The value we measure by subtracting the main cluster component is the luminosity of the subcluster in the case where it is not physically connected to the main cluster. We thus might overestimate the luminosity of the subcluster.
\end{description}

We concentrate in the following on the second case, i.e. correct the estimated luminosity for the remaining main cluster contribution.

We assume axis symmetry around the merger axis of the subcluster. The minor axis --- about 1.4' or 600\,kpc ---, measured in the N-S direction (see also Sec.\ref{morph}) is assumed to be equal to the minor axis of the subcluster perpendicular to the N-S axis and to the merger axis. Furthermore we assume that the subcluster electron density is constant over its volume. If we furthermore assume $d=d_{min}=350$~kpc we can calculate $n_{e,2}$ by integrating the \betam representing the main cluster density distribution (Sec.\ref{sbp}) along the line-of-sight. In this case we find, applying $n_e^2=(n_{e,1}+n_{e,2})^{2}$: $n_{e,2}=$7.1$\times 10^{-4}$\,cm$^{-3}$ for the subcluster electron density.

Since we cannot determine the physical distance d of the subcluster to the main component with high accuracy, we show in Tab.\ref{corlumintab} the luminosity and the mass of the subcluster and the mass ratio between the two components for several values of $d$ ($0.5<d<2$~Mpc).

\begin{table}
\begin{center}
\begin{tabular}{ccccc}
\hline
\hline
Distance (Mpc) & 0.5 & 1 & 1.5 & 2
\\
\hline
$n_{e,1}$ (10$^{-3}$\,cm$^{-4}$) & 10.0 & 2.6 & 1.1 & 0.56
\\
$L_{\sc x}$ (10$^{44}$\,ergs/s) & 0.49 & 1.10 & 1.47 & 1.6
\\
$T$ (keV) & 1.9 & 2.6 & 2.9 & 3.0
\\
$M_{2}$ ($10^{14}\,\mathrm{M}_{\odot}$) & 1.0 & 1.6 & 1.9 & 2.0
\\
$M_{2}/M_{1}$ (\%) & 17 & 27 & 32 & 34
\\
\hline
\end{tabular}
\end{center}
\caption{Corrected luminosity $L_x$ and mass M$_{2}$ of the subcluster depending on the distance of the subcluster to main cluster.
The main cluster density $n_{e,1}$ is calculated according to the \betam presented in Sec.\ref{sbp}. The mass and the temperature T of the subcluster are estimated according to the scaling laws presented in Sec.\ref{totmass}. The mass ratio between main and subcluster is computed assuming for the main cluster component $M_{200}=5.9\times 10^{14} M_\odot$ (see also Sec.\ref{model}).}\label{corlumintab}
\end{table}

Since the subcluster has very likely a physical distance in the order of 0.5--1\,Mpc (Sec.\ref{angle}), we estimate the mass ratio M$_{sub}/$M$_{main}$ to be between 20--30\%. This implies that \object{RX~J0256.5+0006} is a major merger event.

\subsection{The observed subcluster temperature and implications}

One important question is why the region of the subcluster shows such a high temperature. According to the luminosity estimated in Sec.\ref{corlumin} and equ.\,(\ref{lxt}), the highest temperature expected for the subcluster would be about 3.2\,keV. However, we measure a temperature of 5.7$^{+1.3}_{-1.0}$\,keV in this region.

It is clear that the emission of the main cluster contaminates the subcluster region. However, if this was the only contamination, the measured temperature should lie between 3 to 5~keV, but not above 5~keV. This suggests the presence of a third temperature component in this region. This component is quite likely the region in front of the subcluster pointing towards the main cluster centre, which is heated either by adiabatic compression or shock waves due to the infall. The merger geometry, which favors small values for the impact angle $\theta$, maximizes the effect of projection, since the heated region and the subcluster region lie in this case roughly in the same line-of-sight.

In order to estimate the temperature of this third hot component we perform a three temperature fit to the spectrum of the subcluster region. We fix the temperature of the main cluster component (determined from our spectral fits) as well as the subcluster component (determined by converting the observed luminosity of this component in temperature). The intensity of the main cluster component is fixed and calculated from the \betam parameters and the subcluster emission is fixed and estimated from the count rate estimate after main cluster contribution. Unfortunately the statistics of the spectrum is too poor to allow the temperature determination of this third hot component.
 
\section{Discussion and conclusion}\label{conclusion}

We present results of an \xmm observation of \object{RX~J0256.5+0006} a medium distant cluster of galaxies (z=0.36) found in the Bright {\sc sharc} catalog. We observe two X-ray maxima in this cluster. The principal X-ray maximum is linked to the main cluster centre and a second X-ray maximum lies West with respect to the main cluster centre. The observed bimodal morphology of the cluster indicates that this cluster is in a merger state. Hydrodynamic simulations have shown that merging activity in clusters is naturally linked to temperature gradients in the ICM. In our spectro-imaging analysis, however, we do not find strong indications for temperature variations. The lack of temperature gradients is confirmed by spectral fitting in selected regions. 

For the main cluster centre we find a temperature of $kT=4.9^{+0.6}_{-0.5}$~keV, which is representative for the entire main cluster component. We determine the bolo-metric X-ray luminosity of the main cluster to be $L_X=1.5\times 10^{45}$erg/s. This luminosity is higher than the value ($L_X\sim9\times 10^{44}$erg/s) expected from the luminosity-temperature relation established by Arnaud \& Evrard (1999) taking into account redshift evolution (the term: $(1+z)^{3/2}$). Neglecting redshift evolution the discrepancy between observed and expected luminosity becomes even stronger. We take this as an indication that X-ray luminosity evolves with redshift, as predicted from simple theory on structure formation. Our hint for  evolution agrees with a study by Vikhlinin et al. (2002) based on clusters with ($z>0.4$). The high X-ray luminosity of this cluster is probably also linked to the fact that the main cluster in itself is not entirely relaxed. This latter hypothesis is supported by the fact that the surface brightness profiles extracted in different sectors show considerable variation in shape.

X-ray profiles of clusters show a remarkable degree of self-similarity (see for example Neumann \& Arnaud 1999). We compare the X-ray profiles of \object{RX~J0256.5+0006} with the reference profile based on nearby clusters defined in Arnaud, Aghanim \& Neumann (2002). We see that only the surface brightness profile extracted North-East (NE) of the cluster agrees with the reference profile of Arnaud et al. (1999). The other profiles (the overall profile with the subcluster cut out and a profile extracted SE of the cluster) show larger core radii with respect to the reference profile. We take this as indication that the NE sector represents the relaxed part of the cluster. 

Fitting a \betam to the different surface brightness profiles, we observe that $r_c$ (between 100 and 550~kpc) and $\beta$ (0.57-1.2) vary strongly between the different profiles. Consequently the mass estimates based on hydrostatic equilibrium and \betam parameters are quite different. We observe $M_{500}$ between 3.7 and $12\times 10^{14}$\msun. For the NE profile, which seems to represent the relaxed part of the main cluster we find $M_{500}\sim 4\times 10^{14}$ \msun. This is in good agreement with a z-evolving $M-T$ relation, when compared to the results of the HIFLUGCS sample (Finoguenov, Reiprich \& B\"ohringer 2001)  which gives $M_{500}=3.9\times 10^{14}$\msun (transposed into our cosmological model). A non-z-evolving $M-T$ relation gives $M_{500}=6.2\times 10^{14}$\msun, which is only marginally consistent with our observed mass for the main cluster based on the NE surface brightness profile.

With the \betam parameters we also calculate the gas mass of the cluster. The corresponding gas mass fraction ($f_g$) of the cluster varies between 12 and 33\%. Taking into account only the overall and the NE profile we find $f_g$'s in the order of 18--20\%, which is in good agreement with the value found by Arnaud \& Evrard (1999). Pen (1997) suggested that constant $f_g$ in clusters can be used to determine cosmological parameters. Changing cosmology to $\Omega_m=1$, we find indeed lower $f_g$'s around 16--18\%. This is a small indication that indeed $\Omega_m=0.3$ and $\Omega_\Lambda=0.7$, as used in this paper. However, Neumann \& Arnaud (2001) found that the gas mass in clusters does not follow $M_{gas} \propto T^{3/2}$, but $M_{gas} \propto T^{1.94}$. If $M_{tot}\propto T^{3/2}$ as predicted from simple structure formation, or generally $M_{tot}\propto T^{x}$ with $x \neq 1.94$ \footnote{see Horner, Mushotzky \& Scharf (1999), Nevalainen, Markevitch \& Forman (2000) and Finoguenov, Reiprich \& B\"ohringer (2001) for studies on the $M_{tot}-T$ relation}, then $f_g$ varies as function of cluster temperature and indicates that internal cluster physics plays an important role. In this case, using gas mass fractions in clusters for the determination of cosmological parameters is more complicated. One needs to compare $f_g$ of nearby and distant clusters at exactly the same temperature. This approach is furthermore only valid if additionally there does not exist an intrinsic redshift dependence of $f_g$ (see for example Ettori et al. 2004 for the redshift dependence of several physical quantities in clusters). If internal cluster physics change significantly $f_g$ in clusters it is very likely that this effect is redshift dependent. If this is true than using $f_g$ for determining cosmological parameters depends on a large number of physical relationships, which would make this approach less attractive. 

Beside the properties of the main cluster of \object{RX~J0256.5+0006} we also study the cluster's merger dynamics. We develop a simple model to constrain the merger geometry based on an on-axis merger with impact parameter zero. As input we use the projected distance of the subcluster to the main cluster centre observed in X-rays (350~kpc) and estimates of 4 cluster galaxy redshifts obtained with the Kitt Peak telescope. The possible velocity difference between main and subcluster lies between roughly 1800 and 2800~km/s. The range of possible distances ranges between $0.5$~Mpc$ <d<1.4$~Mpc, which limits the impact angle $\theta$ to: $15^\circ<\theta<44^\circ$. The relatively large distances of the subcluster to the main cluster centre are confirmed by the fact that we do not see temperature variations along the merger axis (see above). We also examine the effects of ram pressure stripping in the subcluster ICM. We find that the effects we see in X-rays like cometary shape of the subcluster ICM as well as displacement of X-ray maximum with respect to the main galaxy of the subcluster (galaxy D) suggest distances to the main cluster centre of about 0.6 to 1~Mpc, which imply $20^\circ<\theta<35^\circ$. This is in very good agreement with the simple merger model and suggests that the impact parameter of the merger is very likely close to zero.

Estimating the mass of the subcluster using a luminosity-mass relationship we find that the ratio of subcluster to main cluster mass is in the order of 20 to 30\%. This ratio is relatively high when compared to other cluster mergers and indicates that we observe a major merger in \object{RX~J0256.5+0006}.

\begin{acknowledgements}
We are grateful to Monique Arnaud for useful discussions and for allowing us to use the {\sc Xmm--newton} data prior to publication. We also want to thank Brad Holden and Aronne Merrelli for providing the astrometry of the galaxies presented in this work. We are also grateful to Brad Holden for the optical spectroscopy on the cluster galaxies presented here. We would also like to thank the anonymous referee for useful comments, which improved the paper.

A.K.~R. and R.C.~N. acknowledge partial financial support from the {\sc nasa ltsa} grant {\sc nag5-11634}. D.J.~B. acknowledges the support of {\sc nasa} contract {\sc nas8-39073} ({\sc CXC}).
\end{acknowledgements}

\end{document}